%% file: long-live-blockchain.tex
\newif\iftechrep
\techreptrue

\documentclass[letterpaper,twocolumn,10pt]{article}
\usepackage[section]{placeins} 
\usepackage{graphicx}

\usepackage{enumitem}
\usepackage{amssymb}
\usepackage{amsmath}
\usepackage[outdir=./plots/epstopdf/]{epstopdf}
\graphicspath{{./plots/}}
\usepackage{subcaption}
\usepackage{mathtools} 

\DeclarePairedDelimiter{\ceil}{\lceil}{\rceil}
\usepackage{upgreek}
\usepackage{algorithm}
\usepackage{algorithmicx}
\usepackage{algpseudocode}
\usepackage{amsthm}
\usepackage{macros}
\usepackage{ioa_code_small-1}
\algrenewcommand\textproc{}
\PassOptionsToPackage{hyphens}{url}
\usepackage{hyperref}
\algnewcommand\algorithmicforeach{\textbf{for each}}
\algdef{S}[FOR]{ForEach}[1]{\algorithmicforeach\ #1\ \algorithmicdo}
\usepackage{xspace}
\usepackage[utf8]{inputenc}
\usepackage{textcomp}
\usepackage{booktabs}
\usepackage{tabularx}
\usepackage{multirow}
\usepackage{listings}    
\usepackage{pifont}

\usepackage{subcaption}
\hypersetup{
  colorlinks=true,
  linkcolor=blue!50!blue, 
  urlcolor=blue!70!black,
  citecolor=blue!70!black,
}

\usepackage{ dsfont }
\usepackage{tikz}
\usetikzlibrary{shapes,arrows,positioning,automata,calc}
\tikzset{replica/.style={circle,draw,inner sep=0pt, minimum size=2em}}
\pgfdeclarelayer{background}
\pgfdeclarelayer{foreground}
\pgfsetlayers{background,main,foreground}
\usetikzlibrary{external}

 \usepackage{ wasysym }
\usepackage{xcolor}
\definecolor{olive}{rgb}{0.3, 0.4, .1}
\definecolor{pinegreen}{cmyk}{0.92,0,0.59,0.25}

\usepackage{setspace}

\usepackage{color}
\usepackage{xcolor}
\usepackage[textwidth=0.15\textwidth,textsize=scriptsize]{todonotes}

\usepackage{amsthm}
\usepackage{eufrak}
\newtheorem{definition}{Definition}
\newtheorem{theorem}{Theorem}[section]

\newtheorem{lemma}[theorem]{Lemma}

\newtheorem{innercustomgeneric}{\customgenericname}
\providecommand{\customgenericname}{}
\newcommand{\newcustomtheorem}[2]{%
  \newenvironment{#1}[1]
  {%
   \renewcommand\customgenericname{#2}%
   \renewcommand\theinnercustomgeneric{##1}%
   \innercustomgeneric
  }
  {\endinnercustomgeneric}
}

\newcustomtheorem{customthm}{Theorem}
\newcustomtheorem{customlemma}{Lemma}
\newcustomtheorem{customcor}{Corollary}

\begin{document}
\newcommand{\arcomm}[1]{\todo[color=green,bordercolor=black,linecolor=black]{\textsf{\scriptsize\linespread{1}\selectfont ARP: #1}}}
\newcommand{\arcommin}[1]{\todo[inline,color=green,bordercolor=green,linecolor=green]{\textsf{ARP: #1}}}
\newcommand{\system}{[system name] }

\newcommand{\boxedtext}[1]{\fbox{\scriptsize\bfseries\textsf{#1}}}

\newcommand{\greenremark}[2]{
   \textcolor{pinegreen}{\boxedtext{#1}
      {\small$\blacktriangleright$\emph{\textsl{#2}}$\blacktriangleleft$}
    }}

 \definecolor{burntorange}{rgb}{0.8, 0.33, 0.0}
  \newcommand{\changeremark}[2]{
   \textcolor{burntorange}{\boxedtext{#1}
      {\small$\blacktriangleright$\emph{\textsl{#2}}$\blacktriangleleft$}
}}
\newcommand{\myremark}[2]{
   \textcolor{blue}{\boxedtext{#1}
      {\small$\blacktriangleright$\emph{\textsl{#2}}$\blacktriangleleft$}
    }}
  \newcommand{\myremarknew}[2]{
   \textcolor{violet}{\boxedtext{#1}
      {\small$\blacktriangleright$\emph{\textsl{#2}}$\blacktriangleleft$}
}}
\newcommand{\NewRemark}[2]{
   \textcolor{violet}{
      {\small$\blacktriangleright$\emph{\textsl{#2}}$\blacktriangleleft$}
}}

\newcommand{\redremark}[2]{
   \textcolor{red}{\boxedtext{#1}
      {\small$\blacktriangleright$\emph{\textsl{#2}}$\blacktriangleleft$}
    }}
  
  \newcommand\ARP[1]{\myremark{ARP}{#1}}
  \newcommand\ARPN[1]{\myremarknew{ARPN}{#1}}
\newcommand\vincent[1]{\redremark{VG}{#1}}
\newcommand{\warning}[1]{\redremark{\fontencoding{U}\fontfamily{futs}\selectfont\char 66\relax}{#1}}
\newcommand\NEW[1]{\NewRemark{NEW}{#1}}
\newcommand\UPDATE[1]{\greenremark{TODO}{#1}}
\newcommand\CHANGE[1]{\changeremark{CHANGE(?)}{#1}}


\newcommand{\cref}[1]{{\S\ref{#1}}}
\newcommand{\solution}{ASMR\xspace}
\newcommand{\solutionlong}{Accountable SMR\xspace}
\newcommand{\blockchain}{ZLB\xspace}
\newcommand{\blockchainlong}{Zero-Loss Blockchain\xspace}
\newcommand{\blockchainproblem}{LLB\xspace}
\newcommand{\blockchainlongproblem}{Longlasting Blockchain\xspace}
\newcommand{\component}{BM\xspace}
\newcommand{\componentlong}{Blockchain Manager\xspace}

\newcommand{\statement}{Consider an execution in which an \solution is about to execute a consensus instance  $\Gamma_i$, which tolerates less than $t_0\cdot n$ (i.e. at most $\lceil t_0\cdot n\rceil -1$) failures, where  $t_0\in(0,1)$. }

\newenvironment{smallenum}{
\begin{enumerate}[%
leftmargin=6pt,
labelsep=2pt,
rightmargin=0pt,
labelwidth=2pt,
itemindent=2pt,
listparindent=2pt,
topsep=4pt plus 2pt minus 4pt,
partopsep=4pt,
itemsep=0pt,
parsep=-6pt
]
    \setlength{\parskip}{-1pt}
}{\end{enumerate}}

\newenvironment{smallitem}{
\begin{itemize}[%
leftmargin=6pt,
labelsep=2pt,
rightmargin=0pt,
labelwidth=2pt,
itemindent=2pt,
listparindent=2pt,
topsep=4pt plus 2pt minus 4pt,
partopsep=4pt,
itemsep=0pt,
parsep=-6pt
]
    \setlength{\parskip}{-1pt}
}{\end{itemize}}

\date{}

\title{ZLB: A Blockchain to Tolerate Colluding Majorities}
\author{{\rm Alejandro Ranchal-Pedrosa}\\
  \small{University of Sydney} \\
  \small{Sydney, Australia} \\
  \small{alejandro.ranchalpedrosa@sydney.edu.au}
  \and
  {\rm Vincent Gramoli}\\
  \small{University of Sydney and EPFL}\\
  \small{Sydney, Australia} \\
  \small{vincent.gramoli@sydney.edu.au}}
\maketitle
%


\begin{abstract}
In the general setting, consensus cannot be solved if an adversary controls 
a third of the system. Yet, blockchain participants typically 
reach consensus ``eventually'' despite an adversary controlling 
a minority of the system. Exceeding this $\frac{1}{3}$ cap 
is made possible by tolerating transient disagreements, 
where distinct participants select distinct blocks for 
the same index, before eventually
agreeing to select the same block.
Until now, 
no blockchain could tolerate an attacker controlling a majority of the system.

In this paper, we present \blockchainlong (\emph{\blockchain}), the 
first blockchain that tolerates an adversary controlling more 
than half of the system.
\blockchain is an open blockchain that combines recent theoretical 
advances in accountable Byzantine agreement to 
exclude undeniably deceitful replicas.
Interestingly, \blockchain does not need a known bound 
on the delay of messages but 
progressively reduces the portion of deceitful 
replicas below $\frac{1}{3}$,
and reaches consensus.
Geo-distributed experiments show that \blockchain outperforms HotStuff 
and is almost as fast as the scalable Red Belly Blockchain that cannot tolerate $n/3$ faults.
\end{abstract}


\maketitle



\newcommand{\BVbcast}{\ensuremath{\scriptsize \lit{bv-broadcast}}}
\newcommand{\AUX}{\text{\sc aux}\xspace}
\newcommand{\ECHO}{\text{\sc echo}\xspace}
\newcommand{\EST}{\text{\sc est}\xspace}
\newcommand{\COORD}{\text{\sc coord}\xspace}
\newcommand{\MSG}{\text{\sc msg}\xspace}
\newcommand{\BVAL}{\text{\sc bval}\xspace}
\newcommand{\INIT}{\text{\sc init}\xspace}
\newcommand{\READY}{\text{\sc ready}\xspace}
\newcommand{\CERT}{\text{\sc cert}\xspace}
\algnewcommand{\LeftComment}[1]{\Statex \(\triangleright\) #1}

\input{inc/01introduction.tex}
\input{inc/02background.tex}

\input{inc/03problem.tex}
\input{inc/05solution.tex}
\input{inc/07evaluation.tex}

\input{inc/08related.tex}

\input{inc/09conclusion.tex}\bibliographystyle{abbrv}
\bibliography{long-live-blockchain} 

 \appendix

 \addcontentsline{toc}{section}{Appendices}
\renewcommand{\thesubsection}{\Alph{subsection}}
 \input{inc/appendix}
\end{document}

%% file: inc/01introduction.tex
\section{Introduction}
Blockchain systems~\cite{Nak08} promise to track ownership of assets without a central authority and thus rely heavily on 
distributed nodes agreeing on a unique block at the next index of the chain.
An attacker can exploit a disagreement to \emph{double spend} by simply inserting conflicting transactions in competing blocks.

Some solutions~\cite{BKM18,GAG19,BBC19,CNG21} to this problem avoid forks by
guaranteeing that no 
disagreement can ever occur, even transiently. 
Such solutions typically adopt an open permissioned model where permissionless clients can issue
transactions that $n$ permissioned servers (or \emph{replicas}) encapsulate in blocks they agree upon.
They typically assume partial synchrony~\cite{DLS88} or that there exists an unknown bound on the time it takes to deliver any message.
Unfortunately, it is well-known~\cite{PSL80} that consensus cannot be solved as soon as $\frac{1}{3}$ of these replicas experience a Byzantine fault.
More specifically, in these blockchains an attacker can exploit a disagreement to double spend as soon as it controls $\frac{1}{3}$ of these replicas.

Other solutions, 
popularized by classic blockchains~\cite{Nak08,Woo15}, assume that
the adversary controls only a minority of the replicas, typically expressed as 
computational power or stake.
The tolerance to an adversary controlling more than $\frac{1}{3}$ but less than $\frac{1}{2}$ of the replicas is made possible by accepting forks and tolerating transient disagreements that eventually
get resolved 
in an ``eventual'' consensus.
Unfortunately, as soon as the adversary controls a majority of the system, 
then safety gets violated:
This was recently illustrated by the 
losses of $\$70,000$ and $\$18$ million in Bitcoin Gold~\cite{BCG,BCG2} 
and $\mathdollar 5.6$ million in Ethereum Classic~\cite{ETC}. 

In this paper, we propose the \emph{\blockchainlong} (or \emph{\blockchain} for short), the first blockchain that tolerates more than a majority of failed replicas while assuming partial synchrony. As in classic blockchains~\cite{GKL15,CS20} a client may not be sure that its transaction is already committed but if it manages to send its transaction to an   honest replica, then its transaction is guaranteed to be eventually committed. 
The problem \blockchain solves is not simple for two reasons.

\sloppy{First, if a majority of the permissionned replicas stop participating, we cannot guarantee availability as implied by the CAP theorem~\cite{GilbertCAP}.
To cope with this liveness issue we assume a blockchain-specific failure model called the \emph{deceitful failure model}, similar to 
the alive-but-corrupt failure model~\cite{MNR19} but with weaker assumptions, 
where permissioned replicas are more incentivized to act maliciously than to stop acting forever.
This deceitful failure model departs significantly from the failure models found in closed distributed systems (e.g., datacenters, cloud services or distributed databases) where it is acknowledged~\cite{CKL09,LLM19,KWQ12} that most faults are omissions and rare commissions are due to unlucky events (e.g., disk errors~\cite{CKL09}). 
In open distributed systems, like blockchains,
replicas are incentivized to commit \emph{deceitful faults} 
(sending misinformation to deceive honest replicas)
to double spend and 
long-time \emph{benign faults} (stop sending intelligible messages forever) are typically not as frequent.
Indeed, blockchains are subject to well-fomented attacks to steal assets~\cite{LSP82,EE18} 
but, when not deceitful, blockchain replicas are more likely up and running than crashed forever: they are carefully monitored to generate financial rewards to their owner~\cite{Nak08,Woo15}.} 

Second, if a majority of deceitful replicas collude then they can ensure that honest replicas disagree on the next block.
To cope with this safety issue we rely on recent theoretical advances in the field of accountability~\cite{forensics,CGG21} that guarantee that deceitful replicas leave a cryptographically signed trace at an honest replica when trying to influence the decision of this replica. Honest replicas can then combine these traces to build undeniable proofs of fraud. 
In \blockchain, honest replicas exploit these proofs of fraud to exclude deceitful replicas 
after each disagreement until consensus is reached. One may think that consensus is needed to 
exclude the same replicas.
Interestingly, this is not necessary as, even without knowing when, honest replicas will eventually identify the same set of deceitful replicas, which is finite by 
definition. Thus, after a bounded series of disagreements among $n_0, n_1, \dots, n_k$ replicas, each followed by its corresponding exclusion of more than $n_j/3$ deceitful replicas, the system converges towards $n_k$ with $f_k < n_k/3$ failures, where consensus is reached.

To demonstrate the efficiency of \blockchain we implement it with Bitcoin transactions, compare its performance to modern blockchain systems.
We show that, on 90 machines spread across distinct continents, \blockchain outperforms by $5.6$ times the HotStuff~\cite{YMR19} state machine replication that inspired Facebook Libra's~\cite{BBC19}, and obtains comparable performance to the recent Red Belly Blockchain~\cite{CNG21}. 
Our empirical results also show an interesting phenomenon in that the impact of the attacks decreases rapidly as the system size increases, due to the increased message delays.


As opposed to classic blockchain payment systems~\cite{Nak08,Woo15,LNZ16,EGSvR16} 
that recover from forks a posteriori, 
our payment system guarantees deterministic agreement---no forks---in good executions but recovers only if $f\geq n/3$ by merging decisions.
As opposed to more recent Byzantine fault tolerant solutions~\cite{GHM17,BKM18,GAG19,BBC19,CNG21}, our payment system recovers eventually from a state with
up to either $f<n/3$ Byzantine replicas (the classical scenario), or instead $d<5n/9$ and $3q+d<n$, where $d$ are deceitful replicas and $q$ are benign replicas, being the total number of faulty replicas $f=q+d$. This means for example $f<5n/9$ faulty replicas of which $q<2n/9$ are benign (and the rest deceitful), or $f<2n/3$ and $q<n/6$.


We present the background (\cref{sec:bck}), our \blockchainproblem problem (\cref{sec:pb}), our \blockchain solution (\cref{sec:solution}), our evaluation (\cref{sec:expe})  before 
presenting the related work (\cref{sec:rw}) and concluding (\cref{sec:conclusion}).

%% file: inc/02background.tex
\section{Background and Preliminaries}
\label{sec:bck}
A blockchain system~\cite{Nak08} is a distributed system maintaining a sequence of blocks that contains \emph{valid} (cryptographically signed) and non-conflicting transactions indicating how assets are exchanged between \emph{accounts}.\footnote{Note that in~\cref{sec:solution} we will implement Bitcoin's transactions where ``valid'' implies ``non-conflicting'' as requested transactions cannot be valid if their UTXOs are already  consumed.}
\subsection{Accountability}
The replicas of a blockchain system are, by default, not accountable in that their faults often go undetected. For example, when a replica creates a fork, it manages to double spend after one of the blockchain branches where it spent coins vanishes. This naturally prevents other replicas from detecting frauds and from holding this replica accountable for its misbehavior.
%
%
Recently, Polygraph~\cite{CGG20,CGG21} 
introduced accountable consensus (Def.~\ref{def:accountability})
as the problem of solving consensus if $f<n/3$ and eventually detecting $f_d \geq n/3$ faulty replicas in the case of a disagreement.

\begin{definition}[Accountable Consensus]\label{def:accountability}
The problem of \emph{accountable consensus} is:
\begin{itemize}[leftmargin=*,wide=\parindent]
\item to solve consensus if the number of Byzantine faults is $f<n/3$ and
\item for every honest replica to eventually output at least $f_d= \lceil n/3 \rceil$ faulty replicas if two honest replicas output distinct decisions.
\end{itemize}
\end{definition}

%

\subsection{Byzantine state machine replication} A Byzantine State Machine Replication (SMR)~\cite{CL02,KADC07} is a replicated service that accepts deterministic commands from clients and totally orders these commands using a consensus protocol so that, upon execution of these commands, every honest replica ends up with the same state despite \emph{Byzantine} or malicious replicas. The instances of the consensus execute in sequence, one after the other, starting from index 0. We refer to the consensus instance at index $i$ as $\Gamma_i$.
%

Traditionally, 
given that honest replicas propose a value, the Byzantine consensus problem~\cite{PSL80} is for every honest replica to eventually decide a value (consensus termination), for no two honest replicas to decide different values (agreement) and for the decided value to be one of the proposed values (validity).
In this paper, we consider however a variant of Byzantine consensus (Def.~\ref{def:sbc}) %
useful for blockchains~\cite{MSC16,DRZ18,CNG21} where 
the validity requires the decided value to be a subset of the union of the proposed values, hence allowing us to commit more proposed blocks per consensus instance.
 
\begin{definition}[Set Byzantine Consensus]\label{def:sbc}
Assuming that each honest replica proposes a set of transactions, the \emph{Set Byzantine Consensus} (SBC) problem is for each of them to decide on a set in such a way that the following properties are satisfied:
\begin{itemize}
\item SBC-Termination: every honest replica eventually decides a set of transactions;
\item SBC-Agreement: no two honest replicas decide on different sets of transactions;
\item SBC-Validity: a decided set of transactions is a non-conflicting set of valid transactions taken from the union of the proposed sets; 
\item SBC-Nontriviality:  
if all replicas are honest and propose a common valid non-conflicting set of transactions, then this set is the decided set. 
\end{itemize}
\end{definition}
SBC-Termination and SBC-Agreement properties are common to many Byzantine consensus definition variants, while 
SBC-Validity 
includes two predicates, the first states that 
transactions proposed by Byzantine proposers could be decided as long as they are valid and \emph{non-conflicting} (i.e., they do not withdraw more assets from one account than its balance); the second one is necessary to prevent any trivial algorithm that decides a pre-determined value from solving the problem. 
As a result, we consider that a consensus instance $\Gamma_i$ outputs a set of enumerable decisions $out(\Gamma_i)=d_i,\; |d_i|\in\mathds{N}$ that all  $n$ replicas replicate. We refer to the state of the SMR at the $i$-th consensus instance $\Gamma_i$ as all decisions of all consensus instances up to the $i$-th consensus instance. %



\subsection{Solving the Set Byzantine Consensus (SBC)}
A classic reduction~\cite{BCG93,BKR94,CGLR18,CGG21} of the problem of multi-value consensus, which accepts any ordered set of input values, to the problem of binary consensus, that accepts binary input values, proved promising to solve the SBC problem (Def.~\ref{def:sbc}) when $f<n/3$~\cite{CNG21}.
The idea consists of executing an all-to-all reliable broadcast~\cite{B87} to exchange $n$ proposals: any delivered proposal is stored in an array $\ms{proposals}$ at the index corresponding to the identifier of the broadcaster. A binary consensus at index $k$ is started 
with input value 1 for each index $k$ where a proposal has been recorded. Once $n-f$ proposals are delivered locally, a binary consensus at the remaining indices 
$0\leq \ell < n$ where $\ell\neq k$ is started with input value 0. The results of these concurrent binary consensus instances is stored into a $\ms{bitmask}$ array.
Hence, applying the $\ms{bitmask}$ to the $\ms{proposal}$ array yields a sequence of proposals whose content is the output of consensus. Polygraph is the accountable variant of this algorithm where replicas 
broadcast \emph{certificates}, sets of $2n/3$ messages signed by distinct replicas, each time
they reliably broadcast or decide a binary value.




%% file: inc/03problem.tex
\section{The \blockchainlongproblem Problem}\label{sec:pb}

We considering the classic distributed system model~\cite{CL02,KADC07} 
where messages are delivered within bounded but unknown time (i.e., partial synchrony~\cite{DLS88}) and Byzantine failures are either deceitful or benign. Solving the longlasting blockchain problem is to solve consensus when possible ($f<n/3$), and to recover from a situation where consensus is violated ($n/3 \leq f < 2n/3$) by excluding deceitful replicas to obtain a situation where this violation can be resolved (with $f'<n'/3$).
%

\subsection{\blockchainlongproblem}
A \blockchainlongproblem (\blockchainproblem) is a Byzantine fault tolerant SMR that 
allows for some consensus instances to reach a disagreement before fixing the disagreement by merging the branches of the resulting fork and deciding the union of all the past decisions using SBC (Def.~\ref{def:sbc}). 
More formally, an SMR is an \blockchainproblem if it ensures termination, agreement and convergence:
   
\begin{definition}[Longlasting Blockchain Problem]\label{def:properties} 
%
%
An SMR is an \blockchainproblem if all the following properties are satisfied:
\begin{enumerate}[leftmargin=*,wide=\parindent]
\item {\bf Termination:}
For all $k>0$, consensus instance $\Gamma_k$ terminates, either with agreement or disagreement.
\item {\bf Agreement:}
If $f<n/3$ when $\Gamma_k$ starts, then honest replicas executing $\Gamma_k$ reach agreement.
\item {\bf Convergence:}
There is a finite number of disagreements after which honest replicas agree. 
\end{enumerate}
\end{definition}


Termination does not imply agreement among honest replicas whereas agreement is the classic property of the consensus problem. Convergence preserves the assets of honest replicas by guaranteeing 
that there is a limited number of disagreements (this number is 0 if $f-q<n/3$) after which agreement is maintained, within a sufficiently long static period. 
%
 
\subsection{Threat model}\label{ssec:threat-model}  
We introduce the deceitful failure model that is a refinement of the Byzantine failure model~\cite{LSP82}, radically different from the closed network failure models~\cite{CKL09,KWQ12,LLM19} where most Byzantine failures are omissions and very few are commissions. 
This contrast stems from the observation that blockchain payment systems incentivize replicas to either fail by attacking the network or minimize downtime to maximize profit~\cite{Nak08,Woo15}. 
A \emph{deceitful fault} consists of sending a message that violates the protocol to deceive honest replicas and try to reach a disagreement whereas
a \emph{benign fault} consists of a non-deceitful Byzantine fault (e.g. sending a stale message).
We refer to a replica that commits a deceitful (resp. benign) fault as a \emph{deceitful replica} (resp. \emph{benign replica}).

Note that the deceitful failure model is inspired by the alive-but-corrupt failure model~\cite{MNR19} but relies on a weaker assumption. 
As opposed to the alive-but-corrupt replicas, deceitful replicas do not necessarily know whether their attack will succeed (safety can be violated) before deciding to attack or be honest. Instead, we simply assume that if the attacks of a deceitful replica keep failing then it eventually acts as an honest replica. This assumption is sufficient for our \blockchain to ensure that deceitful replicas failing to violate safety do not prevent termination (Def.~\ref{def:properties}).

We also denote 
the \textit{deceitful ratio} $d/n$ as $\delta$. A replica that is not faulty is \emph{honest}.
Let $n$ be the initial number of replicas in our system, we thus assume a number of faulty replicas that satisfies either $f<n/3$, or $d<5n/9$ and $3q+d<n$, where $d$ are deceitful replicas and $q$ are benign replicas, as the total number of faulty replicas $f=q+d$. 

%
The adversary that controls these faulty replicas is adaptive in
that $f$ can change over time, however, we assume that the adversary is \emph{slowly-adaptive}~\cite{LNZ16}, 
in that the adversary
experiences \emph{static periods} during which deceitful, benign and honest replicas remain so.
We assume that these static periods are long enough for honest replicas to discover and replace the faulty replicas. To model all possible nodes that want to join as replicas in the system we assume that there exists a large pool of $m$ nodes among which at least $2n/3$ are honest nodes ($m$ can be much greater than $n$) and the rest are
deceitful, from which honest replicas will propose to add new
nodes. For ease of exposition, we consider that the pool is of size
$m$ at the beginning of each static period, with at
least $2n/3$ honest nodes, to account for new nodes wanting to join, or recovered nodes previously excluded.

For ease of exposition, we assume a standard
public-key infrastructure (PKI) that associates replicas' identities
with their public-keys, and that is common to all replicas.
We also assume a
computationally bounded adversary as can be found in~\cite{cachin2005random,cachin2001}.

%% file: inc/05solution.tex
\begin{figure}[t]
  \includegraphics[scale=0.5]{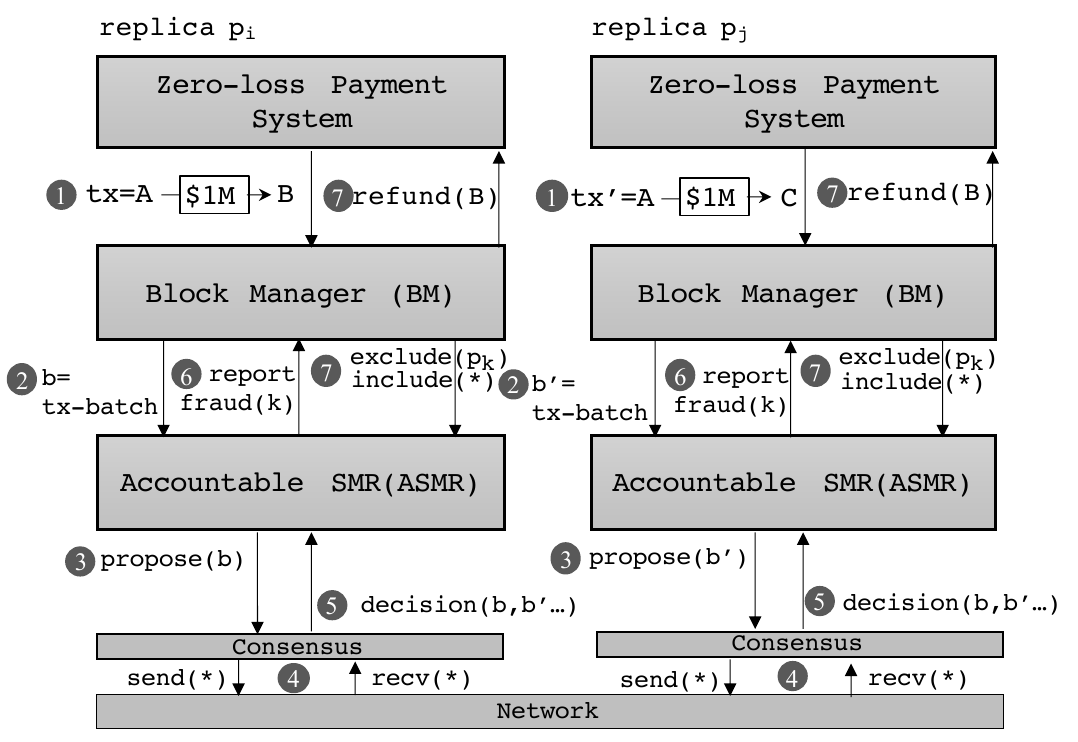} 
\caption{The distributed architecture of our \blockchain system relies on \solution, \component and the payment system 
deployed on several replicas. {\color{darkgray}\ding{203}} Each replica batches some payment requests illustrated with {\color{darkgray}\ding{202}} a transfer $\ms{tx}$ (resp. $\ms{tx}'$) of \$1M from Alice's account (A) to Bob's (B) (resp. Carol's (C)). Consider that Alice has \$1M initially and attempts to double spend by modifying the code of a replica $p_k$ under her control so as to execute a coalition attack. 
{\color{darkgray}\ding{204}--\ding{206}} The \solution component detects the deceitful replica $p_k$ that tried to double spend, the associated transactions $\ms{tx}$ and $\ms{tx'}$ and account $A$ with insufficient funds. It chooses transaction $\ms{tx}$ and discards $\ms{tx}'$, {\color{darkgray}\ding{207}} notifies \component that {\color{darkgray}\ding{208}} excludes or replaces replica $p_k$ and {\color{darkgray}\ding{208}} refunds $B$ with $p_k$'s deposit.\label{fig:architecture}
}
\vspace{-1.5em}
\end{figure}
\section{The \blockchainlong}\label{sec:solution}
In this section we detail our system. 
Its two main ideas are (i)~to replace deceitful replicas undeniably responsible for a fork by new replicas to converge towards a state where consensus can be reached, and (ii)~to refund conflicting transactions that were wrongly decided. 
%
We will show that \blockchain solves the \blockchainlongproblem. 
%
As depicted in Figure~\ref{fig:architecture}, we present below the components of our system, namely the \solution (\cref{ssec:asmr}) and the \componentlong(\component) (\cref{ssec:blockchain})
\iftechrep
 but we defer the zero loss payment application
(\cref{ssec:payment}).
\fi

\begin{figure*}[t]
\begin{center}
\includegraphics[scale=0.54]{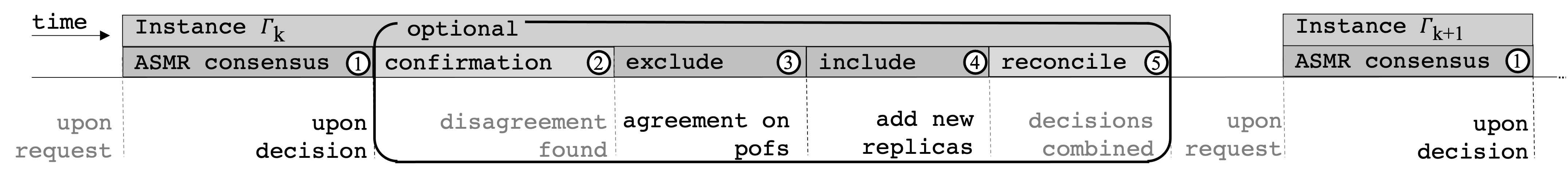}
\caption{If there are enqueued requests that wait to be served, then a replica starts a new instance $\Gamma_k$ by participating in an \solution consensus phase \ding{192}; a series of optional phases may follow: \ding{193}~the replica tries to confirm this decision to make sure no other honest replica disagrees, \ding{194}~it invokes an exclusion consensus if it finds enough proofs of fraud (PoFs), \ding{195}~it then potentially includes new replicas to compensate for the exclusion, and \ding{196}~merges the two batches of decided transactions. Some of these phases complete upon consensus termination (in black) whereas other phases terminate upon simple notification reception (in grey). The replica starts a new instance $\Gamma_{k+1}$ if there are other enqueued requests to be served, hence participating in a new \solution consensus phase \ding{192} that may succeed, in which case none of the optional phases immediately follow.\label{fig:phases}
}
\end{center}
  \vspace{-2em}
\end{figure*}

As long as new requests are submitted by a client to a replica, the payment system component of the replica converts it into a payment that is passed to the \component component. As depicted in Fig.~\ref{fig:architecture}, when sufficiently many 
payment requests have been received, the \component issues a batch of requests to the \solution that, in turn, proposes 
it to the consensus component. The consensus component exchanges messages through the network for honest replicas to agree. If a disagreement is detected, then the account of the deceitful replica is slashed. 
Consider that Alice (A) attempts to double spend by (i)~spending her \$1M with 
both Bob (B) and Carol (C) in $\ms{tx}$ and $\ms{tx}'$, respectively, and (ii)~hacking the code of replica $p_k$ that commits deceitful faults to produce a disagreement. Once the \solution{} detects the disagreement, \component is notified, 
replica $p_k$ is excluded or replaced and $B$ is refunded with $p_k$'s deposit.

\subsection{\solutionlong (\solution)}\label{ssec:asmr}

In order to detect deceitful replicas, we
now present, as far as we know, the first accountable state machine replication, called \solution.
\solution consists of running an infinite sequence of five actions: \ding{192}~the accountable consensus (Def.~\ref{def:accountability}) that tries to decide upon a new set of transactions, \ding{193}~a confirmation that aims at confirming that the agreement was reached, \ding{194}--\ding{195}~a membership change that aims at replacing deceitful replicas responsible for a disagreement by new replicas and \ding{196}~a reconciliation phase that combines all the decisions of the disagreement, as depicted in Figure~\ref{fig:phases}.

\subsubsection{The phases of \solution}
For each index, \solution first executes the accountable consensus (\cref{sec:bck}) phase \ding{192} to try to agree on a set of transactions then it optionally runs four subsequent phases \ding{193}--\ding{196} to recover from a possible disagreement.
\begin{enumerate}[leftmargin=* ,wide=\parindent]
\item[\ding{192}] {\bf \solution consensus:} Honest replicas propose a set of transactions, which they received from clients, to the accountable consensus (Def.~\ref{def:accountability}) in the hope to reach agreement. When the consensus terminates, 
all honest replicas agree on the same decision or some honest replicas disagree: they decide distinct sets of transactions. 
\item[\ding{193}] {\bf Confirmation:} As honest replicas could be unaware of the other decisions, 
they enter an optional confirmation phase waiting for messages coming from more distinct replicas than what consensus requires. 
If the confirmation leads honest replicas
to detect 
disagreements, 
i.e., they receive certificates supporting distinct decisions, 
then they start the membership change. 
This phase may not terminate 
as an honest replica needs to deliver messages from more than $(\delta + 1/3) \cdot n$ 
replicas (due to the number of `conflicting histories'~\cite{singh2009zeno})
to guarantee that no disagreement was possible by a deceitful ratio $\delta$,
however, it does not prevent $\Gamma_k$ from terminating as  $\Gamma_k$ proceeds in parallel with the confirmation without waiting for its termination. If the confirmation phase terminates, it either confirms that a block is irrevocably final (no replica disagreed), or a membership change starts.

\item[\ding{194}-\ding{195}]\textbf{Membership change:} 
Our membership change (Alg.~\ref{alg:recconsensus}) consists of two 
consecutive consensus algorithms: one that excludes deceitful replicas (line~\ref{line:slashing-consensus-1}), another that adds newly joined replicas (line~\ref{line:slashing-consensus-2}).
We
separate inclusion and exclusion in two consensus instances to avoid
deciding to exclude and include replicas proposed by the same
replica. 
Replica $p_i$ maintains a series of
variables: the current consensus instance $\Gamma_k$, the
$\ms{deceitful}$ replicas among the whole set $C$  of current replica
ids, a set of replica ids that is updated at runtime for the exclusion consensus $C'$, the pool of replicas
$\ms{pool}$, a set of certificates $\ms{certificates}$, a set of
proofs of fraud (PoFs) $\ms{pofs}$ and of new PoFs
$\ms{new\_pofs}$,
 a local threshold $f_d$ of detected deceitful replicas, a
set $\ms{cons-exclude}$ of decided PoFs and a set $\ms{cons-include}$ of
decided new replicas.
\begin{enumerate}[leftmargin=* ,wide=2\parindent]
\item[\ding{194}] \textbf{Exclusion consensus:}
Honest replicas identify deceitful replicas by cross-checking their certificates. These cross-checks produce undeniable PoFs with conflicting signed messages from the same replica, indicating equivocation.
If honest replicas detect $f_d= \lceil n/3 \rceil$ deceitful replicas
(via distinct PoFs), they stop their pending ASMR consensus (line~\ref{line:instance-paused}) before restarting it with the new set of replicas (line~\ref{line:instance-restarted}). Then, honest replicas start the membership change ignoring
messages from these $f_d$ replicas by using instead an updated committee $C'$ that
already discards these replicas (lines~\ref{line:discardstart}-\ref{line:slashing-consensus-1}). Honest replicas propose their set of PoFs at the start of the exclusion consensus $\lit{ex-propose}$ by invoking the Polygraph~\cite{CGG21} accountable consensus algorithm (line~\ref{line:slashing-consensus-1}). 

The key novelty of our algorithm is for replicas to exclude other replicas, and thus update their committee $C'$, at runtime upon reception of new valid PoFs (lines~\ref{line:upcom1}-\ref{line:upcom2}).
To this end, honest replicas validate a delivered $\lit{ex-propose}$ certificate (line~\ref{line:checkcert2}), and decide the same decision with the same associated certificate if it contains $\ceil{2|C'|/3}$ signatures from replicas that are not discarded (line~\ref{line:certdec}).
As $C'$ decreases at runtime, this threshold is guaranteed to be eventually met.
Upon updating their committee, honest replicas re-check all their certificates (line~\ref{line:checkcert1}) and re-broadcast their PoFs (line~\ref{line:brpofs}).
\iftechrep
(See Lemma~\ref{lemma:recovery} for the proof of correctness.)
\fi
As our exclusion consensus solves SBC (Definition~\ref{def:sbc}), it maximizes the number of excluded replicas by deciding at least $\ceil{2|C'|/3}$ proposals at once, 
where $C'$ is the number of replicas in the committee when the exclusion consensus terminates.

Note that instead of waiting for $f_d$ PoFs (Line~\ref{line:if-ex-propose-not-start}), 
the replicas could start Alg.~\ref{alg:recconsensus} as soon as they detect one deceitful replica, however, waiting for at least $f_d$ PoFs guarantees that a membership change is necessary and will help remove many deceitful replicas from the same coalition at once. Moreover, waiting for $f_d$ PoFs allows us to guarantee agreement of the exclusion consensus 
\iftechrep
in Lemma~\ref{lemma:recovery}.
\fi 
\item[\ding{195}] \textbf{Inclusion consensus:}
To compensate with the excluded replicas, an inclusion protocol (line~\ref{line:slashing-consensus-2}) adds new candidate replicas taken from the pool of candidates (\cref{ssec:threat-model}) in line~\ref{line:pool}.
This inclusion protocol is also an instance of Polygraph, like the exclusion protocol, except it differs in the format and verification of the proposals, each containing as many new replicas to include as replicas were excluded by the exclusion consensus (lines~\ref{line:pool}-\ref{line:slashing-consensus-2}). Also, the inclusion consensus does not update its committee at runtime, and instead it uses the updated committee, where $C$ already excluded at least $f_d$ replicas (line~\ref{line:exclude}).
Since the union of the $\ceil{2|C|/3}$ proposals contain more than enough replicas to include, we apply a deterministic function $\lit{choose}$ (line~\ref{lin:deter}) to the union of all decided proposals, such that all included replicas are selected as evenly as possible from all decided proposals, in order to restore the original committee size to $n$. This guarantees (i)~a fair distribution of inclusions across all decisions, and (ii)~that the membership change does not increase the deceitful ratio even if all included replicas are deceitful. At the end, the excluded replicas are punished by the application layer (line~\ref{line:punish}) and the new replicas are included (lines~\ref{lin:deter}-\ref{line:instance-restarted}) to compensate for the discarded replicas.

Honest replicas accept certificates (in consensus instances
that follow the membership change) from non-excluded replicas
containing signatures from excluded replicas. This is because honest replicas from different partitions of a disagreement might not find themselves at the same consensus instance $\Gamma_k$ at the time that they execute the membership change.
Notice that this occurs during a transient period of time
since all valid certificates contain at least $1$ honest
replica by construction, and all honest replicas eventually update their committee and stop generating new certificates with excluded replicas. This transient acceptance allows honest
replicas to catch up without reverting any decision.
\end{enumerate}
\item[\ding{196}] {\bf Reconciliation:} Upon delivering a conflicting block with an associated valid certificate, the reconciliation starts by combining all transactions that were decided by distinct honest replicas in the disagreement. These transactions are ordered through a deterministic function, whose simple example is a lexicographical order but can be made fair by rotating over the indices of the instances.
\end{enumerate}

Once the current instance $\Gamma_k$ terminates, another instance $\Gamma_{k+1}$ can start, even if it runs concurrently with a confirmation or a reconciliation at index $k$ or at a lower index.

    \begin{algorithm}[!htbp]
      \caption{Membership change at replica $p_i$ and consensus $\Gamma_k$}
      \label{alg:recconsensus}
      \smallsize{
        \begin{algorithmic}[1]
          \Part{\smallsize {\bf State}}{
            \State $\Gamma_k$, $k^{\ms{th}}$ instance of ASMR consensus $p_i$ participates to.
            \State $C$, set of replicas forming the committee
            \State $C'$, updated set of replicas, initially $C'=C$
            \State $\ms{certificates}$, received certificates during exclusion, initially $\emptyset$
            \State $\ms{pofs}$, the set of proofs of fraud (PoFs), initially $\emptyset$
            \State $\ms{new\_pofs}$, set of newly delivered PoFs, initially $\emptyset$
            \State $\ms{cons-exclude}$, the set of PoFs output by consensus, initially $\emptyset$
            \State $\ms{cons-include}$, the set of new replicas output by consensus, initially $\emptyset$
            \State $\ms{pool}$, the pool of replicas from which to propose new replicas
            \State $\ms{deceitful}\in I$, the identity of an agreed deceitful replica, initially $\emptyset$
            \State $f_d$, the threshold of proofs of fraud to recover, $\lceil n/3 \rceil$ by default
        }\EndPart
        
        \Statex \rule{0.45\textwidth}{0.4pt}
        \Part{\smallsize \textbf{Upon receiving a list of proofs of fraud} $\ms{\_pofs}$}{
          \SmallIf{$\lit{verify(}\_\ms{pofs}\lit{)}$}{} \label{line:verify-pof}\Comment{if PoFs are correctly signed}
          \State $\ms{new\_pofs}\gets \_pofs\backslash pofs$
          \State $\ms{pofs}.\lit{add(}\_\ms{pofs}{)}$ \Comment{add PoFs on distinct replicas}
          \SmallIf{$\lit{ex-propose}$ \textbf{not started}}{} \label{line:if-ex-propose-not-start}
            \SmallIf{$\lit{size(}\ms{pofs}\lit{)}\geq f_d$}{} \label{line:pof-size}\Comment{enough to change members}
            \SmallIf{$\Gamma_k$ started \textbf{and not} finished}{}
            $\Gamma_k$.\lit{stop}() \label{line:instance-paused}
            \EndSmallIf
            \State $C'\gets C'\backslash \ms{new\_pofs}.\lit{replicas()}$ \label{line:discardstart}
            \State $\lit{ex-propose.update\_committee(}C'\lit{)}$\Comment{update committee}
            \State $\lit{ex-propose.start(}\ms{pofs}\lit{)}$ \label{line:slashing-consensus-1}\Comment{exclusion consensus}
            
            \EndSmallIf
            \EndSmallIf
            \SmallElseIf{$\ms{new\_pofs}\neq \emptyset$ \textbf{and} $\lit{ex-propose}$ \textbf{not} finished} \label{line:upcom1}
            \State $C'\gets C'\backslash \ms{new\_pofs}.\lit{replicas()}$
            \State $\lit{ex-propose.update\_committee(}C'\lit{)}$\Comment{update committee}\label{line:upcom2}
            \State $\lit{broadcast(}\ms{new\_pofs}\lit{)}$\Comment{broadcast new PoFs}\label{line:brpofs}
            \State $\lit{ex-propose.check\_certificates(}\ms{certificates}\lit{)}$\Comment{recheck certificates}\label{line:checkcert1}
            \EndSmallElseIf
            
            \EndSmallIf}\EndPart
          \Statex \rule{0.45\textwidth}{0.4pt}
          \Part{\smallsize \textbf{Upon receiving a certificate of the exclusion consensus} $\ms{ex-cert}$}{
            \SmallIf{$\ms{ex-cert}\not\in\ms{certificates}$ \textbf{and} $\lit{verify\_certificate}(\ms{ex-cert})$}\label{line:upcom3}
              \State $\ms{certificates.add(}\ms{ex-cert}\lit{)}$
                
                \EndSmallIf
              \State $\lit{ex-propose.check\_certificates(}\{\ms{ex-cert}\}\lit{)}$\Comment{check certificate with current $C'$}\label{line:checkcert2}
            }\EndPart
            \Statex \rule{0.45\textwidth}{0.4pt}
            \Part{\textbf{function} $\lit{ex-propose.check\_certificates(}\ms{certs}\lit{)}$}{
              \For{all $\ms{cert}\in\ms{certs}$}
              \SmallIf{$\lit{verify\_certificate(}\ms{cert}\lit{)}$}{}
                \SmallIf{$|\ms{cert}\lit{.replicas()}\cap C'|\geq \frac{2|C'|}{3}$}{} \Comment{current threshold}
                \State $\lit{ex-propose.cert\_decide(}\ms{cert}\lit{)}$ \Comment{decide certificate's decision}\label{line:certdec}
                \EndSmallIf
                
              \EndSmallIf
                
              \EndFor

            }\EndPart
            \Statex \rule{0.45\textwidth}{0.4pt}
            \Part{\textbf{Upon deciding a list of proofs of fraud $\ms{cons-exclude}$ in $\lit{ex-propose}$}}{
                        	\State $\lit{detected-fraud(}\ms{cons\_exclude}.\lit{get\_replicas()}\lit{)}$\label{line:punish} \Comment{application  punishment} 
			\State $\ms{pofs}\gets \ms{pofs} \setminus \ms{cons-exclude.}\lit{get\_pofs()}$ \Comment{discard the treated pofs}
			\State $\ms{C}\gets \ms{C} \setminus \ms{cons-exclude}.\lit{get\_deceitfuls()}$ \Comment{exclude deceitful}\label{line:exclude}
                        \State $\ms{inc-prop}\gets \ms{pool}\lit{.take}(|\ms{cons-exclude}|)$\Comment{take replicas from the pool}\label{line:pool}
                        \State $\lit{inc-propose.start(}\ms{inc-prop}\lit{)}$ \label{line:slashing-consensus-2}\Comment{inclusion cons.}

                      }\EndPart
                        \Statex \rule{0.45\textwidth}{0.4pt}
                        
                        \Part{Upon deciding a list of replicas to include $\ms{cons-include}$ in $\ms{inc-propose}$}{
                          \State $\ms{new\_replicas}\gets \lit{choose(}|\ms{cons-exclude}|\lit{,}\ms{cons-include}\lit{)}$\Comment{deterministic}\label{lin:deter}
                        \For{all $\ms{new\_replica}\in\ms{new\_replicas}$}  \Comment{for all new to inc.}
                        \State $\lit{set-up-connection}(\ms{new\_replica})$ \Comment{new replica joins}
                        \State $\lit{send-catchup}(\ms{new\_replica})$\label{line:cu}\Comment{get latest state}
                        \EndFor
                        \State $C\gets C\cup \ms{new\_replicas}$
	 \SmallIf{$\Gamma_k$ stopped}{}
              {\bf goto} \ding{192} of Fig.~\ref{fig:phases} \label{line:instance-restarted} \Comment{restart cons.}
              \EndSmallIf
              }\EndPart

          \end{algorithmic}
          }
        \end{algorithm}

\subsection{\componentlong (\component)}\label{ssec:blockchain}

We now present the \componentlong (\component) 
that builds upon \solution to merge the blocks from multiple branches of a blockchain when forks are detected. 
%
Once a fork is identified, the conflicting blocks are not discarded as it would be the case in classic blockchains when a double spending occurs, but they are merged. Upon merging blocks, \component also copes with conflicting transactions, as the ones of a payment system, by taking the funds of excluded replicas to fund conflicting transactions.
\iftechrep
We defer to~\cref{ssec:payment} 
\fi
the details of the amount replicas must have on a deposit to guarantee this funding.


Similarly to Bitcoin~\cite{Nak08}, 
\component accepts transaction requests from a permissionless set of users.
In particular, this allows users to use different devices or wallets to issue distinct transactions withdrawing from the same account---a feature that is not offered in payment systems without consensus~\cite{CGK20}.
In contrast with Bitcoin, but similarly to recent blockchains~\cite{GHM17,CNG21}, our system does not incentivize all users to take part in 
trying to decide upon every block, instead a restricted set of permissioned replicas have this responsibility for a given block. This is why \component, combined with \solution, offers what is often called an open permissioned blockchain~\cite{CNG21}. Nevertheless, \solution can offer a permissionless blockchain with committee sortition without substantial modifications.

\subsubsection{Guaranteeing consistency across replicas}
Building upon the accountability of the underlying \solution that resolves disagreement, \component
features a block merge to resolve forks by excluding deceitful replicas and including new replicas.
%
A consensus may reach a disagreement  if $f\geq n/3$, resulting in the creation of multiple branches or blockchain forks. 
%
%
\component builds upon the membership change of \solution in order to recover from forks. 
In particular, the fact that \solution excludes
$f_d$ deceitful replicas each time a disagreement occurs guarantees that  
the ratio of deceitful replicas $\delta = d/n$ converges to a state where consensus is guaranteed%
\iftechrep
 (Lemma~\ref{lemma:recovery}). 
\fi
The maximum number of branches that can result from forks depends on the number $q$ of benign faults and the number $d$ of deceitful faults as was already 
shown for histories of SMRs~\cite{singh2009zeno}.

       \begin{algorithm}[ht]
      \caption{Block merge at replica $p_i$
      } \label{alg:blkmerg}
     \smallskip
     \smallsize{
      \begin{algorithmic}[1]

		\Part{{\bf \smallsize State}}{
			\State $\Omega$, a blockchain record with fields: 
			\State \T$\ms{deposit}$, an integer, initially $0$  \label{line:deposit}
			\State \T$\ms{inputs-deposit}$, a set of deposit inputs, initially in the first deposit \label{line:inputs-deposit}
			\State \T$\ms{punished-acts}$, a set of punished account addresses, initially $\emptyset$ \label{line:punished-accounts}
			\State \T$\ms{txs}$, a set of UTXO transaction records, initially in the genesis block 
			\State \T $\ms{utxos}$, a list of unspent outputs, initially in the genesis block
		}\EndPart
		
		\Statex \rule{0.45\textwidth}{0.4pt}
		
		\Part{\smallsize Upon receiving conflicting block $\ms{block}$} { \label{line:safe-binpropose} \Comment{merge block}
			 \For{$\ms{tx}$ \textbf{in} $\ms{block}$} \Comment{go through all txs}
        				\SmallIf{$\ms{tx}$ \textbf{not in} $\Omega.\ms{txs}$}{} \Comment{check inclusion}
                                        \State $\lit{CommitTxMerge(\ms{tx})}$ \label{line:merge-invocation} \Comment{merge tx, go to line~\ref{line:merge}}
                                 \For{$\ms{out}$ \textbf{in} $\ms{tx.outputs}$}\Comment{go through all outputs}
       				\SmallIf{$\ms{out.account}$ \textbf{in} $\Omega.\ms{punished-acts}$}{} \Comment{if punished}
        				\State $\lit{PunishAccount(\ms{out.account})}$ \Comment{punish also this new output}
        				\EndSmallIf
                                        \EndFor
        				\EndSmallIf
        			\EndFor
       			\State $\lit{RefundInputs()}$  \Comment{refill deposit, go to line~\ref{line:refund}}
        			\State $\lit{StoreBlock(\ms{block})}$ \Comment{write block in blockchain}
		}\EndPart

      		\Statex  \rule{0.45\textwidth}{0.4pt}
		
		\Part{\smallsize $\lit{CommitTxMerge(\ms{tx})}$}{ \label{line:merge}
        	 		\State $\ms{toFund} \gets 0$
        			\For{$\ms{input}$ \textbf{in} $\ms{tx.inputs}$}\Comment{go through all inputs}
        				\SmallIf{$\ms{input}$ \textbf{not in} $\Omega.\ms{utxos}$}{}\Comment{not spendable, need to use deposit}
        					\State $\Omega.\ms{inputs-deposit}.\lit{add(\ms{input})}$ \Comment{use deposit to refund}
        					\State $\Omega.\ms{deposit}\gets \Omega.\ms{deposit}-\ms{input.value}$\label{line:deposit-withdrawal} \Comment{deposit decreases in value}
        				\EndSmallIf
        				\SmallElse{} $\Omega.\lit{consumeUTXO(\ms{input})}$\Comment{spendable, normal case}
        				\EndSmallElse
        			\EndFor\label{line:merge-end}
		}\EndPart
		
      		\Statex \rule{0.45\textwidth}{0.4pt}
		
		\Part{ \smallsize $\lit{RefundInputs()}$}{ \label{line:refund}
        			\For{$\ms{input}$ \textbf{in} $\Omega.\ms{inputs-deposit}$}\Comment{go through inputs that used deposit}
        				\SmallIf{$input$ \textbf{in} $\Omega.\ms{utxos}$}{}\Comment{if they are now spendable}
        					\State $\Omega.\lit{consumeUTXO(}input\lit{)}$\Comment{consume them}
        					\State $\Omega.\ms{deposit}\gets \Omega.\ms{deposit}+\ms{input.value}$\Comment{and refill deposit}
        				\EndSmallIf       
			\EndFor
		}\EndPart
		
      \end{algorithmic}
		}
    \end{algorithm}

\subsubsection{In memory transactions}\label{ssec:utxo}
\blockchain is a blockchain that inherits the same \emph{Unspent Transaction Output (UTXO)} model of Bitcoin~\cite{Nak08};
the balance of each account in the system is stored in the form of a UTXO table. 
In contrast with Bitcoin, the number of maintained UTXOs is kept to a minimum in order to allow in-memory optimizations. Each entry in this table is a UTXO that indicates some amount of coins that a particular account, the `output' has. When a transaction transferring from source accounts $s_1, ..., s_x$ to recipient accounts $r_1, ..., r_y$ executes, it checks the UTXOs of accounts $s_1, ..., s_x$. If the UTXO amounts for these accounts are sufficient, then this execution consumes as many UTXOs as possible and produces another series of UTXOs now outputting the transferred amounts to $r_1, ..., r_y$ as well as what is potentially left to the source accounts $s_1, ..., s_x$. Maximizing the number of UTXOs to consume helps keeping the table compact.
Each replica can typically access the UTXO table directly in memory for faster execution of transactions.

\subsubsection{Protocol to merge blocks} As depicted in Alg.~\ref{alg:blkmerg}, the state of the blockchain $\Omega$ consists of a set of inputs $\ms{inputs-deposit}$ (line~\ref{line:inputs-deposit}), a set of account addresses $\ms{punished-acts}$ (line~\ref{line:punished-accounts}) that have been used by deceitful replicas, a $\ms{deposit}$ (line~\ref{line:deposit}), 
that is used by the protocol, a set $\ms{txs}$ of transactions and a list $\ms{utxos}$ of UTXOs. 
The algorithm propagates blocks by broadcasting on the network and starts upon reception of a block that conflicts with an existing known block of the blockhain $\Omega$ by trying to merge all transactions of the received block with the transactions of the blockchain $\Omega$ (line~\ref{line:merge-invocation}). This is done by invoking the function $\lit{CommitTxMerge}$ (lines~\ref{line:merge}--\ref{line:merge-end}) where the inputs get appended to the UTXO table and conflicting inputs are refunded with the deposit (line~\ref{line:deposit-withdrawal}) of deceitful replicas. 
\iftechrep
We explain in~\cref{ssec:payment} 
\fi
how to build a payment system with a sufficient deposit 
to remedy double spending attempts.


\subsubsection{Cryptographic techniques}
To provide authentication and integrity, the transactions are signed using the Elliptic Curves Digital Signature Algorithm (ECDSA) with parameters \texttt{secp256k1}, as in Bitcoin~\cite{Nak08}. Each honest replica assigns a strictly monotonically increasing sequence number to its transactions.
The network communications use gRPC between clients and replicas and raw TCP sockets between replicas, but all communication channels are encrypted through SSL. Finally, the excluding protocol (Alg.~\ref{alg:recconsensus}) also makes use of ECDSA for PoFs or authenticating the sender of messages responsible for disagreement. One may think that message authentication codes (MACs) or threshold encryption could be more efficient alternatives to this classic public-key cryptosystem, however, threshold encryption cannot be used to trace back the faulty users as they are encoded in less bits than what is needed to differentiate users, and MACs are insufficient to provide this transferrable authentication~\cite{CJK12}.

%% file: inc/07evaluation.tex
\section{Experimental Evaluation}\label{sec:expe}
%
This section is dedicated to answer the following questions: Does \blockchain offer practical performance in a geo-distributed environment?
When $f<n/3$ how does \solution perform compared to the raw  state machine replication at the heart of Facebook Libra~\cite{BBC19} and the recent fast Red Belly Blockchain~\cite{CNG21}?
When Libra and Red Belly Blockchain are unsafe (i.e., $f\geq n/3$), what is the impact of large scale coalition attacks on the recovery of \solution?  
\iftechrep
We defer the evaluation of a zero-loss payment application to~\cref{ssec:eval-payment}. 
\fi

\paragraph{Selecting the right blockchains for comparison} 
As we offer a solution for open networks, we cannot rely on the synchrony assumption made by 
other blockchains~\cite{GHM17}. 
As we need to reach consensus, we have to assume an unknown bound on the delay of messages~\cite{DLS88}, and do not compare against randomized blockchains~\cite{MSC16,DRZ18,GLT20,LZT20} whose termination is yet to be proven~\cite{TG19}.
This is why, we focus our evaluation on partially synchronous blockchains.
We thus evaluated Facebook Libra~\cite{BBC19}, however, its performance were limited to 11 transactions per second, seemingly due to its Move VM overhead, we thus omit these results here and focus on its raw state machine replication (SMR) algorithm, HotStuff and its available C++ code that was previously shown to lower communication complexity of traditional Byzantine fault tolerance SMRs~\cite{YMR19}.
We also evaluate the blockchain based on Polygraph~\cite{CGG21} as it is, as far as we know, the only blockchain that relies on an accountable consensus protocol. Even though it detects deceitful failures, this blockchain does not tolerate more than $n/3$ failures as it cannot recover after detection. Finally, we evaluate the recent scalable Red Belly Blockchain~\cite{CNG21} as it is one of the fastest blockchains to date.

\paragraph{Geodistributed experimental settings}
We deploy the four systems 
in two distributed settings of c4.xlarge Amazon Web Services (AWS) instances equipped with 4 vCPU and 7.5\,GiB of memory: (i)~a LAN with up to 100 machines 
and (ii)~a WAN with up to 90 machines.
We evaluate \blockchain with a number of failures $f$ up to $\lceil \frac{2n}{3} \rceil - 1$, however, when not specified we fix $d=\lceil 5n/9\rceil-1$ and $q=0$.
All error bars represent the 95\% confidence intervals and the plotted values are averaged over 3 to 5 runs.  
All transactions are \textasciitilde{}400-byte Bitcoin transactions with ECDSA signatures~\cite{Nak08}.

\begin{figure}[t]
  \hspace{-1.4em}
  \includegraphics[height=6.5em]{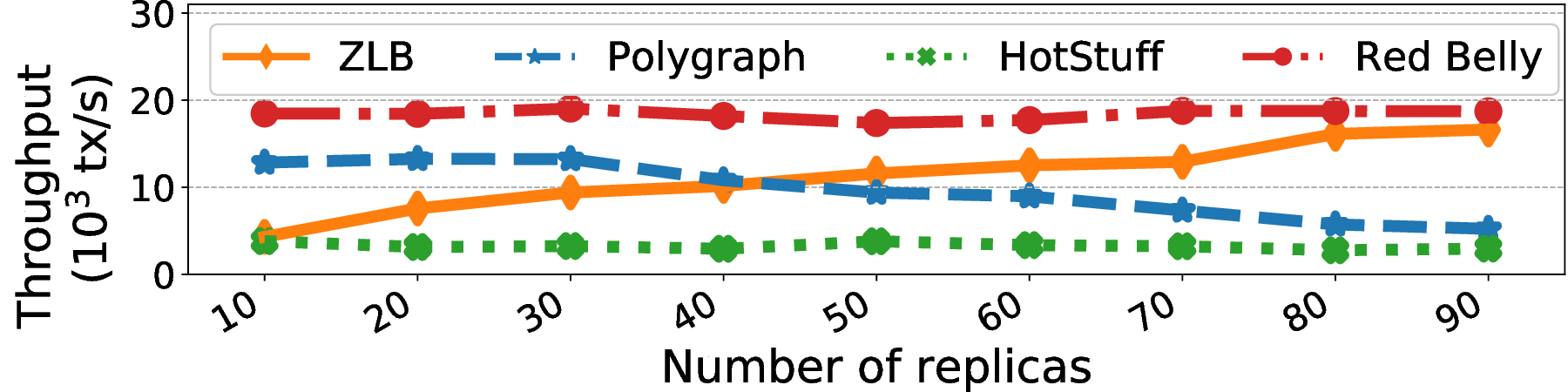}
  \caption{Throughput of \blockchain 
  compared to that of Polygraph~\cite{CGG21}, HotStuff~\cite{YMR19} and Red Belly Blockchain~\cite{CNG21}.} 
  \label{fig:fig1}
  \vspace{-1.5em}
\end{figure}

\subsection{\blockchain vs. HotStuff, Red Belly and Polygraph}
Figure~\ref{fig:fig1} compares the performance of \blockchain, Red Belly Blockchain and Polygraph deployed over 5 availability zones of 2 continents, California, Oregon, Ohio, Frankfurt and Ireland (exactly like Polygraph experiments~\cite{CGG21}).
For \blockchain, we only represent the decision throughput that reaches $16,626$ tx/sec at $n=90$ as the confirmation throughput is similar ($16,492$ tx/sec). As only \blockchain tolerates $f\geq n/3$, we fix $f=0$. 

First, Red Belly Blockchain offers the highest throughput.
As expected it outperforms \blockchain due to its lack of accountability: it does not require messages to piggyback certificates to detect PoFs. 
Both solutions solve SBC so that they decide more transactions as the number of proposals enlarges and use the same batch size of $10,000$ transactions per proposal.
As a result \solution scales pretty well: the cost of tolerating $f\geq n/3$ failures even appears negligible at $90$ replicas.

Second, HotStuff offers the lowest throughput even if it does not verify transactions. 
Note that HotStuff is benchmarked with its dedicated clients in their default configuration, they transmit the proposal to all servers to save bandwidth by having servers exchanging only a digest of each transaction. 
The performance is explained by the fact that HotStuff decides one proposal per consensus instance, regardless of the number of submitted transactions, which is confirmed by previous observations~\cite{VG19}. 
By contrast, \blockchain becomes faster as $n$ increases to outperform HotStuff by $5.6\times$ at $n=90$.

Finally, Polygraph is faster at small scale than \blockchain, because Polygraph's distributed verification and reliable broadcast implementations~\cite{CGG21} are not accountable, performing less verifications. 
After 40 nodes, Polygraph becomes slower than \blockchain because of our optimizations: e.g., its RSA verifications are larger than our ECDSA signatures and consume more bandwidth. 

\begin{figure}[t]
 \hspace{-1.5em}
  \includegraphics[height=10.5em]{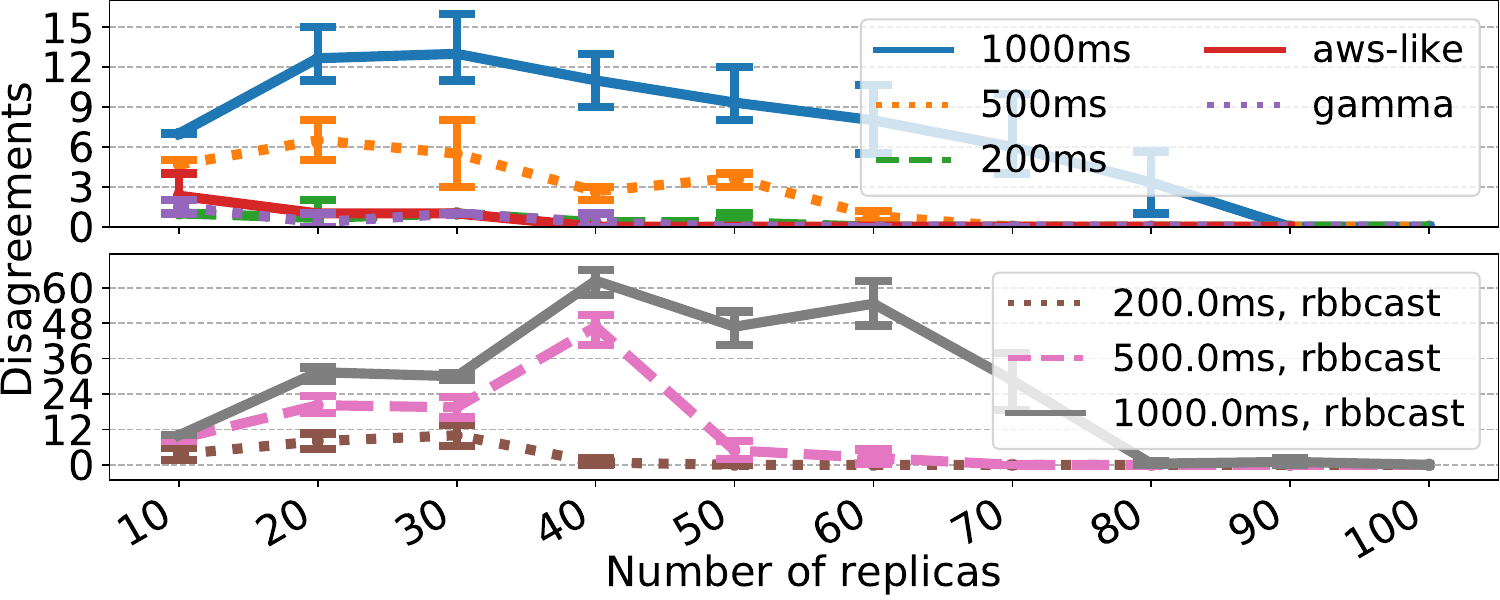}
  \caption{Disagreeing decisions per number of replicas for various uniform delays and for delays generated from a Gamma distribution and a distribution that draws from observed AWS latencies, when equivocating while voting for a decision (top), and while broadcasting the proposals (bottom), for $f=\lceil 5n/9\rceil-1$ and $q=0$.}
  \label{fig:fig2}
  \vspace{-1.5em}
\end{figure}

\subsection{Scalability of \blockchain despite coalition attacks}
To evaluate \blockchain under failures, we implemented both coalition attacks: the reliable broadcast attack and the binary consensus attack (\cref{sec:pb}) with
$d=\lceil 5n/9\rceil-1$ deceitful replicas and $q=0$ benign replicas.
To disrupt communications between partitions of honest replicas, we inject random communication delays between partitions based on the uniform and Gamma distributions, and the AWS delays obtained in previously published measurements traces~\cite{mukherjee1992dynamics,crovella1995dynamic,CNG21}.   
(Deceitful replicas communicate normally with each partition.) 

Fig.~\ref{fig:fig2}(top) depicts the amount of disagreements as the number of distinct proposals decided by honest replicas, as triggered by the binary consensus attack.
First, we select uniformly distributed delays between the two partitions with mean as high as 200, 500 and 1000 milliseconds.
Then, we select delays following a Gamma distribution with parameters taken from~\cite{mukherjee1992dynamics,crovella1995dynamic} and a distribution that randomly samples the fixed latencies previously measured between AWS regions~\cite{CNG21}. 
We automatically calculate the maximum amount of branches that the size of deceitful faults can create (i.e., 3 branches for $d<5n/9$), we then create one partition of honest replicas for each branch, and we apply these delays between any pair of partitions.

Interestingly, we observe that our agreement property is scalable: the greater the number of replicas (for the same relative deceitful ratio), the harder for attackers to cause disagreements. This interesting scalability phenomenon is due to an unavoidable increase of the communication latency between attackers as the scale enlarges, which gives relatively more time for the partitions of honest replicas to detect the deceitful replicas that have equivocated and construct PoFs, hence limiting the number of disagreements.

%
With more realistic network delays (Gamma distribution and AWS latencies) that are lower in expectation than the uniform delays,  
deceitful replicas can barely generate a single disagreement, let alone with an increasing number of replicas.  
This confirms the scalability of our system.

Fig.~\ref{fig:fig2}(bottom) depicts the amount of disagreements under the reliable broadcast attack. Recall that this attack consists of having deceitful replicas equivocating when expected to reliably broadcast the same proposal, hence sending different proposals to different partitions. The number of disagreements is substantially higher during this attack than during the previous attack, however, it drops faster as the system enlarges, 
because the attackers expose themselves earlier.
\begin{figure*}[t]
      \begin{subfigure}{0.495\columnwidth}
        \includegraphics[width=.93\textwidth]{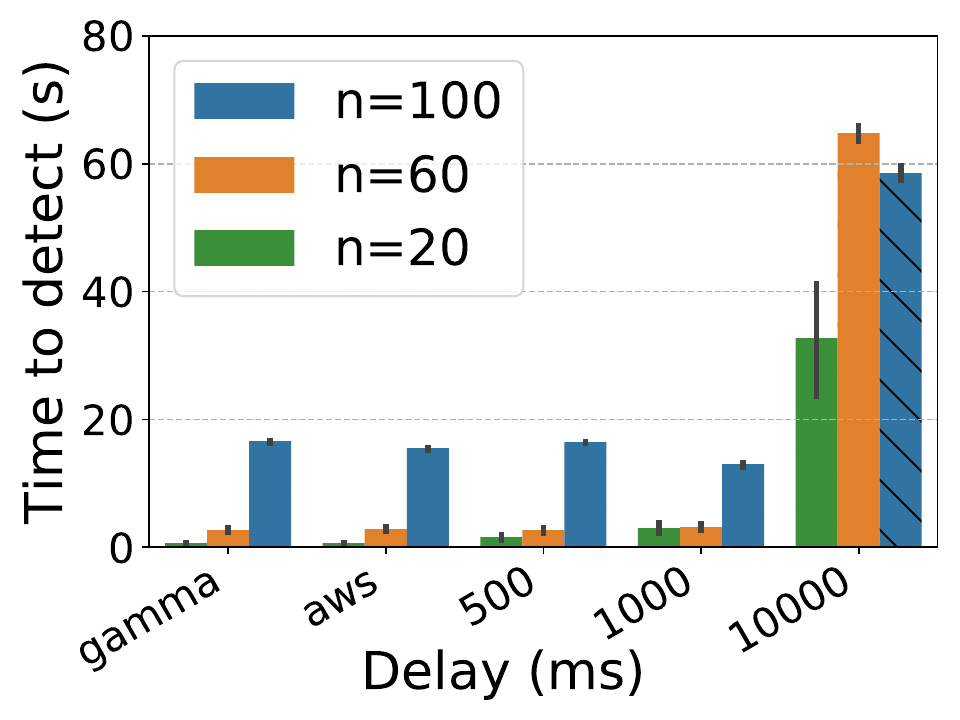}
      \end{subfigure}
      \begin{subfigure}{0.495\columnwidth}
        \includegraphics[width=.93\textwidth]{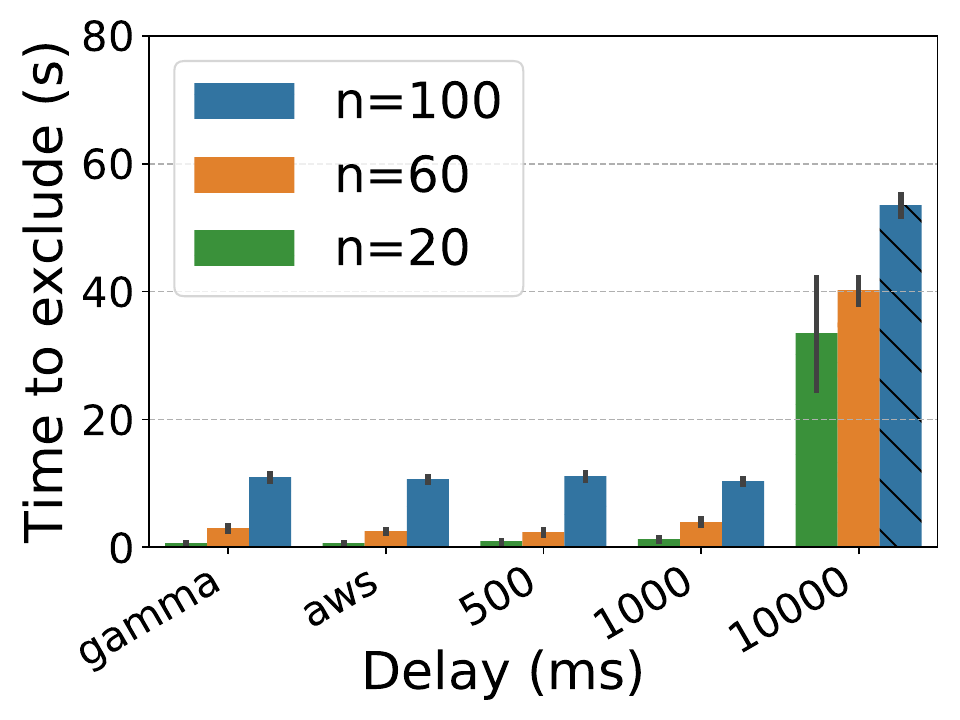}
      \end{subfigure}
      \begin{subfigure}{0.495\columnwidth}
        \includegraphics[width=.93\textwidth]{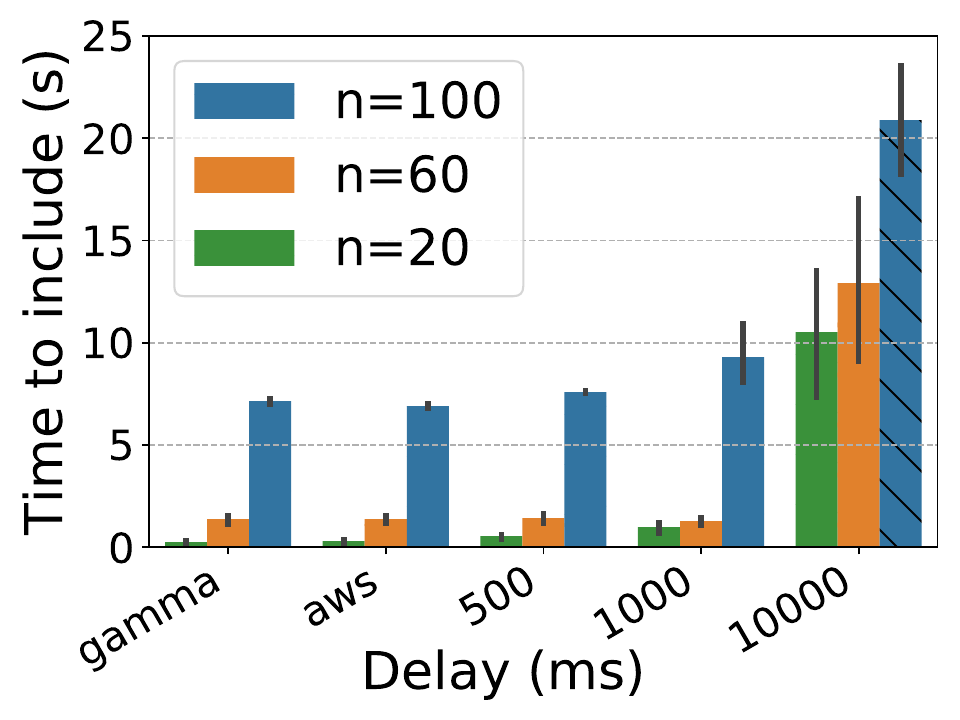}
      \end{subfigure}
      \begin{subfigure}{0.495\columnwidth}
        \includegraphics[width=.93\textwidth]{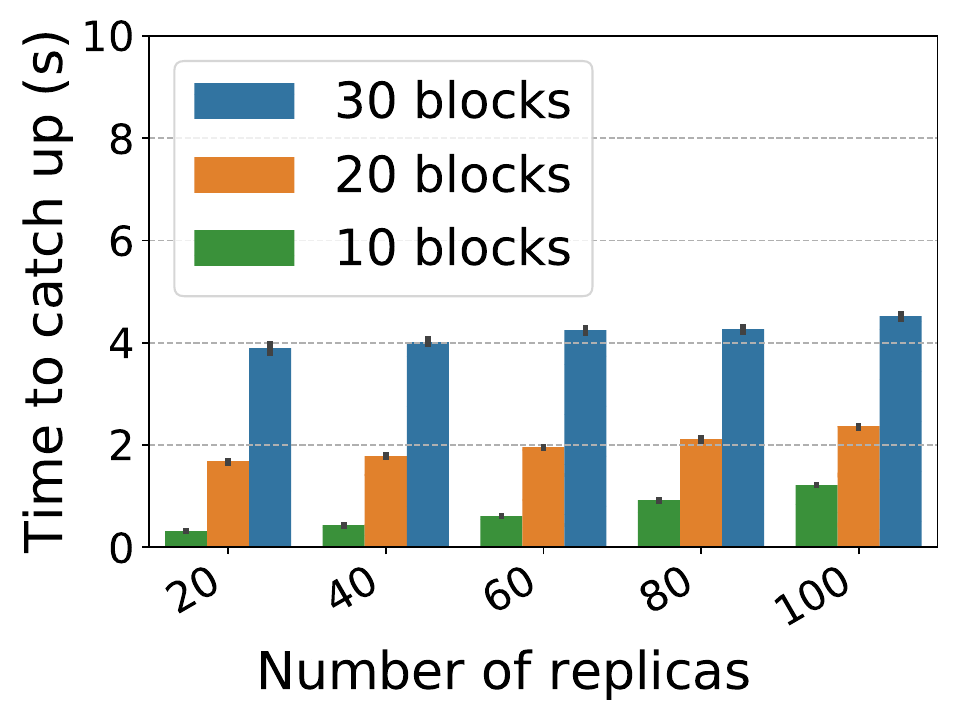}
      \end{subfigure}
      \caption{(Left to right) Time to detect $\ceil{\frac{n}{3}}$ deceitful replicas, exclude them, include new replicas, per delay distribution and number of replicas; and catch up per number of blocks and replicas, with $f=\lceil 5n/9\rceil-1$.}
      \label{fig:fig7}
  \vspace{-1em}
    \end{figure*}
    \subsection{Disagreements due to failures and delays}
We now evaluate the impact of even larger coalitions and delays on \blockchain, we measure the number of disagreements as we increase the deceitful ratios and the partition delays in a system from 20 to 100 replicas. 
Note that these delays could be theoretically achieved with man-in-the-middle attacks, but are notoriously difficult on real blockchains due to direct peering between the autonomous systems of mining pools~\cite{EGJ18}.


While \blockchain is quite resilient to attacks for realistic but not catastrophic delays, attackers can wait and try to attack when the network collapses for a few seconds between regions. Our experiments also show result for the the amount of proposals attackers can disagree on in such a catastrophic scenario, reaching up to 52 disagreeing proposals for a uniform delay of 10 seconds between partitions of honest replicas for the binary consensus attack, and up to 33 disagreements for a uniform delay of 5 seconds, with $n=100$. Further tests showed that the reliable broadcast attack reaches up to 165 disagreeing proposals with a 5-second uniform delay.
\subsection{Time to merge blocks and change members} 
To have a deeper understanding of the cause of \blockchain delays, we measured the time needed 
to merge blocks and to change the membership by replacing deceitful replicas by new ones.

\begin{table}[h]
\normalsize{
  \setlength{\tabcolsep}{17pt}
  \centering
  \begin{tabular}{l|rrr}
  \toprule
Blocksize (txs)& 100  & 1000 & 10000 \\
Time (ms) & 0.55
& 4.20
& 41.38
\\
  \bottomrule
\end{tabular}
}
\setlength{\tabcolsep}{20pt}
  \caption{Time to merge locally two blocks for different sizes with all transactions conflicting.} 
      \label{tab:01}
      \vspace{-1em}
    \end{table}

Table~\ref{tab:01} shows the times to locally merge two blocks for different sizes assuming the worst case: all transactions conflict. This is the time taken in the worst case because replicas can merge proposals that they receive concurrently (i.e., without halting consensus).
It is clear that this time to merge blocks locally is negligible compared to the time it takes to run the consensus algorithm.

Figure~\ref{fig:fig7} shows the time to detect $f_d \geq \ceil{\frac{n}{3}}$ deceitful replicas (left), to run the exclusion consensus (center-left), and to run the inclusion consensus (center-right), for a variety of delays and numbers of replicas. The time to detect reflects the time from the start of the attack until honest replicas detect the attack:
If the first $\ceil{\frac{n}{3}}$ deceitful replicas are forming a coalition together and creating the maximum amount of branches, then the times to detect the first deceitful replica and the first $\ceil{\frac{n}{3}}$ deceitful replicas overlap entirely, as they are cross-checking conflicting certificates.
(We detect all at the same time.) 

Moreover, the times to detect, exclude, and include increase as the communication delays increase. The time to exclude (57 seconds) is significantly larger than to include (21 seconds), due to the proposals of the exclusion consensus carrying PoFs and leading replicas to execute a time consuming cryptographic verification. With shorter communication delays, performance becomes practical.
Finally, Figure~\ref{fig:fig7} (right) depicts the time to catch up depending on the number of proposals (i.e., blocks). As expected, this time increases linearly with the number of replicas, due to the catchup requiring to verify larger certificates, but it remains nevertheless practical at $n=100$ nodes.

%% file: inc/08related.tex
\section{Related Work}\label{sec:rw}
Several works tried to circumvent the upper bound on the number of Byzantine failures~\cite{LSP82} to reach agreement.
As opposed to permissionless blockchains~\cite{Nak08}, open permissioned blockchains
try to rotate the consensus participants to cope with an increasing amount of colluding replicas 
without perfect synchrony~\cite{VG19b,BAS20}. The notion of accountability has originally been applied to distributed systems in PeerReview~\cite{HKD07} and to consensus protocols in Polygraph~\cite{CGG20}, however, not to recover from inconsistencies.

Slashing stakes aims at disincentivizing blockchain participants to misbehave. The Casper~\cite{Casper} algorithm incurs a penalty in case of double votes but does not ensure termination
when $f<n/3$ as one replica can always restart the consensus instance. 
Tendermint~\cite{BKM18} aims at slashing replicas, but is not accountable.
Balance~\cite{HGG19} adjusts the size of the deposit to avoid over collateralizing but we are not aware of any system that implements it.
SUNDR~\cite{LKM04} 
assumes honest clients that communicate directly to detect Byzantine failures.
Polygraph~\cite{CGG21} solves the accountable consensus but does not provide slashing. FairLedger~\cite{levari2019fairledger} assumes synchrony in order to detect faulty replicas. Sheng et al.~\cite{forensics} consider forensics support as the ability to make replicas accountable for their actions to clients. We do not consider this model in which they show accountability cannot be achieved with $2n/3$ faults.
%
Freitas de Souza et al.~\cite{SKRP21} reconfigure replicas in a lattice agreement after detection.
Shamis et al.~\cite{shamis2021pac} propose, in a concurrent work with ours, 
to store signed messages in a dedicated ledger so as to punish replicas 
in case of misbheavior. As it builds upon PBFT it needs more than $2n/3$ honest 
replicas to guarantee progress. 
 
Traditionally, closed distributed systems consider that omission faults (omitting messages) are more frequent than commission faults (sending wrong messages)~\cite{CKL09,KWQ12,LLM19}.
Zeno~\cite{singh2009zeno} guarantees eventual consistency by decoupling requests into weak (i.e., requests that may suffer reordering) and strong requests. It provides availability in periods where $f$ goes beyond $n/3$ by committing only weak requests, that can be reordered later. 
\blockchain could not be built upon Zeno because Zeno requires wrongly ordered transactions to be rolled back, whereas blockchain transactions can have irrevocable side effects like the shipping of goods to the buyer.
BFT2F~\cite{Li} offers fork* consistency, which forces the adversary to keep correct clients in one fork, while also allowing accountability. Stewart et al.~\cite{grandpa} provide a finality gadget similar to our confirmation phase, however, it does not recover from disagreements.
The BAR model~\cite{AAC05}, and its Byzantine, altruistic, rational classification 
is motivated by multiple administrative domains and 
corresponds better to the blockchain open networks without distinguishing benign from deceitful faults.

Various non-blockchain systems already refined the types of failures to strengthen guarantees.
Upright~\cite{CKL09} proposes a system that supports $n=2u+r+1$ faults, where $u$ and $r$ are the numbers of commission and omission faults, respectively. They can either tolerate $n/3$ commission faults or $n/2$ omission faults.
Flexible BFT~\cite{MNR19} offers a failure model and theoretical results to support $\lceil 2n/3\rceil -1$ alive-but-corrupt replicas. An alive-but-corrupt replica only behaves maliciously when it can violate safety, but it behaves correctly otherwise. This assumption is too strong for our needs: for example, it is difficult for a replica to detect whether its coalition is large enough for its attack to succeed.

Some hybrid failure models tolerate crash failures and Byzantine
failures but prevent Byzantine failures from partitioning the
network~\cite{LVC16}. Others aim at guaranteeing that well-behaved
quorums are responsive~\cite{LLM19} or combine crash-recovery with
Byzantine behaviors to implement reliable broadcast~\cite{BC03}. We
are not aware of any deceitful failure model. 
Neu et
al.~\cite{ebbnflow} show an implementation of a system that provides
availability in partial synchrony for $f<n/3$ and finality in
synchrony for $f<n/2$. However, their system does not tolerate
$n/3$ deceitful failures in partial synchrony. They also
extend this work to identify an accountability-availability dilemma that motivates the
need for deceitful replicas~\cite{neu2021availability}.

%% file: inc/09conclusion.tex
\section{Conclusion}\label{sec:conclusion}

In this paper, we proposed \blockchain, the first blockchain that tolerates an adversary controlling a majority of failures by exploiting accountability and with a new deceitful failure model.
\blockchain outperforms HotStuff and achieves close 
performance to the recent Red Belly Blockchain. 

%% file: inc/appendix.tex
\subsection{Fault tolerance and proofs}\label{sec:proof-sketch}
In this section, we show that \blockchain solves the \blockchainlongproblem problem (Def.~\ref{def:properties}) hence recovering from of a majority of failures.
To this end, we show that \blockchain solves termination and agreement (when $f<n/3$), and convergence. 

A membership change is called \emph{successful} if its exclusion and inclusion consensus lead to a state in which all honest replicas decide on the same set of replicas to exclude and to include.
\begin{lemma}

In \blockchain, the exclusion consensus solves consensus.
   \label{lemma:exc}
 \end{lemma}
 \begin{proof}[Proof]

   Validity is immediate from the fact that honest replicas only decide at least
   $f_d\geq n/3$ valid PoFs proposed by another replica. We consider now agreement and termination.

\noindent {\bf Agreement.}  
     By construction, honest replicas only start the membership change if they gather at least $f_d\geq n/3$ PoFs from distinct replicas, which later they propose during the exclusion consensus. As a result, we obtain agreement of the exclusion consensus, because $(d-f_d)<(n-f_d)/3$, and thus $d<5n/9$, regardless of whether honest replicas initially discard the same or different sets of at least $f_d$ faulty replicas.

 \noindent {\bf Termination.} Note that $q\leq f-f_d$, since $f_d\geq n/3$ must be deceitful for the membership change to even start. Deceitful replicas either help terminating or generate conflicting messages by definition. If they generate conflicting messages, they will be removed from the exclusion consensus as the committee updates at runtime. We consider then the most difficult scenario for termination of the membership change, in which all deceitful replicas behaved deceitful signing conflicting messages. Notice also that replicas modify the committee at the start of the exclusion consensus by already discarding the replicas for which they hold PoFs. This means that different replicas can discard different sets of faulty replicas during the exclusion consensus. While this can prevent termination temporarily, since all replicas will eventually receive the same set of PoFs, and since they rebroadcast their updated sets of PoFs and re-check received certificates, they will eventually discard the same set of replicas and update their committee size for the exclusion consensus. As such, all honest replicas either terminate or eventually receive all PoFs excluding all deceitful replicas while executing the exclusion consensus, leading to a committee size of $n'=n-d$. Since termination requires $q<n'/3$, we have that $q<(n-d)/3$ which means $3q+d<n$.

We have thus that the exclusion consensus guarantees termination and agreement, satisfying consensus.\end{proof}

\begin{lemma} 
  In \blockchain, the inclusion consensus solves consensus.
  \label{lemma:inc}
  \end{lemma}
  \begin{proof}[Proof]
    The result is analogous to that of the exclusion consensus in Lemma~\ref{lemma:exc}, and thanks to the deterministic function common to all replicas, with the added simplicity that all honest replicas agree on the set of the excluded replicas at the beginning of the inclusion consensus (because the inclusion consensus executes after the exclusion consensus and the exclusion consensus satisfies agreement , as we showed in Lemma~\ref{lemma:exc}).
  \end{proof}
\begin{lemma} 
In \blockchain, 
any disagreement is followed by a membership change that terminates successfully, and that removes at least $f_d\geq n/3$ replicas. 
   \label{lemma:recovery}
 \end{lemma}
 \begin{proof}
       We consider the case $f\geq n/3$ since for $f<n/3$ the membership change does not even start. Recall that the number of benign replicas $q$ satisfies $3q+d<n$. Note that $d\leq f<\frac{2n}{3}$ implies deceitful replicas cannot generate a valid certificate on their own.
Since honest replicas exchange certificates following the Polygraph protocol~\cite{CGG21},
they 
identify at least $f_d\geq n/3$ deceitful replicas responsible for a disagreement.\\

if $d<n/3$ then $q<2n/9$ satisfying termination and also agreement because $d<n/3$, meaning the membership change does not even start. For the cases where the membership change starts (i.e. $\frac{n}{3}\leq d$), lemmas~\ref{lemma:exc} and~\ref{lemma:inc} show that the exclusion and inclusion satisfy consensus.
   \end{proof}

 Recall that by assumption (\cref{sec:pb})
 there exists a pool with $2n/3$ honest replicas and $m-2n/3$ deceitful replicas, from which honest replicas select a subset to propose to the inclusion consensus. 
 For simplicity, we also assume that no replica from this pool is included twice.
 \begin{theorem}[Convergence] 
In \blockchain, for a sufficiently long static period of the slowly-adaptive adversary,
there is a finite number of disagreements after which honest replicas agree.
   \label{theorem:convergence}
 \end{theorem}
 \begin{proof}[Proof]
 If $d<n/3$ there is no disagreement, and because $q<2n/9$ there is also termination. If instead $d\geq n/3$, there are at least $f_d$ deceitful replicas, which implies that there will be termination of this iteration either with or without a disagreement.
 By Lemma~\ref{lemma:recovery}, we know that every disagreement leads to a membership change in which $f_d \geq n/3$ will be removed.
The inclusion consensus does not increase the deceitful ratio, 
 since the inclusion consensus does not include more replicas than the number of excluded replicas by the exclusion consensus (thanks to the deterministic function) and all excluded replicas are deceitful. 
 As the inclusion consensus decides at least $2n'/3$ proposals where $n'\in[n-f_d,n-d]$ and the remaining deceitful replicas are 
 $d'\leq d-f_d<n/3< 2n'/3$, it follows that some proposals from honest replicas will be decided. As the pool of joining candidates is finite and no replica is included more than once, it follows that the deceitful ratio will eventually decrease in any sufficiently long static period of the adversary, after enough membership changes add enough honest replicas from the pool.
Some inclusion consensus will thus eventually lead to a deceitful ratio $\delta < 1/3$ (i.e. $d<n/3$) and agreement is reached from then on, since $q<n/3$ as well.
\end{proof}

Agreement and termination follows from the correctness of the underlying consensus~\cite{CGLR18,TG19}.
 In particular, it relies on the existing proof that it solves SBC (Def.~\ref{def:sbc})~\cite{CNG21} and the existing formal verification of its binary consensus~\cite{BGK21}.


\subsection{A Zero-Loss Payment Application}\label{ssec:payment}
\label{ssec:eval-payment}

In this section, we describe how \blockchain can be used to implement a \emph{zero-loss payment system} where no honest replica loses any coin.  
The key idea is to request the consensus replicas to deposit a sufficient amount of coins in order to spend, in case of an attack, the coins of deceitful replicas to avoid any honest replica loss.

 \paragraph{Attacking the SBC solution} In the solution to the SBC problem (Def.~\ref{def:sbc}), deceitful replicas can form a coalition of $f\geq n/3$ replicas to lead honest replicas to a disagreement by \emph{equivocating} (sending distinct messages) to different partitions of honest replicas, with one of two coalition attacks:
\begin{enumerate}[leftmargin=*,wide=\parindent]
\item {\bf Reliable broadcast attack:} deceitful replicas misbehave during the reliable broadcast by sending different proposals to different partitions, leading honest replicas to end up with distinct proposals at the same index $k$. For example, deceitful replicas send block $b_a$ with transaction $tx_a$ to a subset $A$ of honest replicas, while block $b_b$ with conflicting transaction $tx_b$ to a subset $B$ of honest replicas, $A\cap B=\emptyset$, both at the same index $k$. 
\item  {\bf Binary consensus attack:}  deceitful replicas vote for each binary value in each of two partitions for the same binary consensus leading honest replicas to decide different bits in the same index of their bitmask, where deciding $1$ (resp. 0) at bitmask index $k$ means to include (resp. not include) proposal at index $k$ in the Blockchain. For example, deceitful replicas send messages to decide $1$ and $0$ to a subset of honest replicas $A$, while they send messages to decide $0$ and $1$ to a subset $B$ of honest replicas, with $A\cap B=\emptyset$, on the binary consensus instances associated to block $b_a$ with transaction $tx_a$ and block $b_b$ with conflicting transaction $tx_b$, respectively.
\end{enumerate}

Note that deceitful replicas do not benefit from combining these attacks: If two honest replicas deliver different proposals at index $k$, the disagreement comes from them outputting 1 at the corresponding binary consensus instance. Similarly, forcing two honest replicas to disagree during the $k^{\ms{th}}$ binary consensus only makes sense if they both have the same corresponding proposal at index $k$.

 \paragraph{Assumptions}
In order to measure the expected impact of a coalition attack succeeding with probability $\rho$ 
in forking \blockchain by leading a consensus to disagreement, we first need to make the following assumptions:
\begin{enumerate}[leftmargin=*,wide=\parindent] 
 \item {\bf Fungible assets.}
We assume that users can transfer assets (like coins) that are \emph{fungible} in that one unit is interchangeable and indistinguishable from another of the same value. An example of a fungible asset is a cryptocurrency.
\item {\bf Deposit refund.}
  To limit the impact of one successful double spending on a block, \blockchain keeps the deposit for a number of blocks $m$, before returning it. A transaction should not be considered \textit{final} (i.e. irreversible) until it reaches this blockdepth $m$. We call thus $m$ the \textit{finalization blockdepth}. Attackers can fork into $a$ branches, and try to spend multiple times an amount $\mathfrak{G}$ (per block), which we refer to as the \textit{gain}, obtaining a maximum gain of $(a-1)\mathfrak{G}$. 
  Each correct replica can calculate the gain by summing up all the outputs of all transactions in their decided block. Additionally, replicas can limit the gain to an upper-bound by design, discarding blocks whose sum of outputs exceed the bound, or they can allow the gain to be as much as the entire circulating supply of assets. The \textit{deposit} $\mathfrak{D}$ is a factor of the gain, i.e., $\mathfrak{D}= b\cdot \mathfrak{G}$. The goal is for every coalition to have at least $\mathfrak{D}$ deposited, and since every coalition has at least size $\ceil{n/3}$, this means that each replica must deposit an amount $3b\mathfrak{G}/n$.

%
 \item {\bf Network control restriction.}
Once faulty replicas select the disjoint subsets (i.e., the partitions) of honest replicas to double spend, 
we need to prevent Byzantine replicas from communicating infinitely faster than honest replicas in different partitions.
%
More formally, let $X_1$ (resp. $X_2$) be the random variables that indicate the time it takes for a message between two replicas within the same partition (resp. two honest replicas from different partitions). We have $E(X_1) / E(X_2) > \varepsilon$, for some $\varepsilon > 0$. Note that the definition of $X_1$ also implies that it is the random variable of the communication time of either two honest replicas of the same partition or two Byzantine replicas. Notice that this probabilistic synchrony assumption is similar to that of Bitcoin and other blockchains that guarantee exponentially fast convergence, a result that also holds for \blockchain under the same assumptions. Notwithstanding, we show in the following an analysis focusing on the attack on each consensus iteration, considering a successful disagreement if there is a fork in a single consensus instance, even for a short period of time. 
\end{enumerate}

\paragraph{Theoretical analysis.}
We show that attackers always fund at least as much as they steal.
We consider that a membership change starts before a disagreement occurs or does not start, which is safer than the general case. Therefore, the attack represents a Bernoulli trial that succeeds with probability $\rho$ (per block) that can be derived from $\varepsilon$.
Out of one attack attempt, the attackers may gain up to $(a-1)\mathfrak{G}$ coins by forking or lose at least $\mathfrak{D}$ coins 
as a punishment from the system. In case of a successful attack, the system compensates the coin losses using the deposited coins.

We introduce the random variable $Y$ that measures the number of
attempts for an attack to succeed and follows a geometric
distribution with mean $E(Y)=\frac{1-\hat{\rho}}{\hat{\rho}}$, where
$\hat{\rho} = 1 - \rho$ is the probability that the attack fails. Thus, we
define the expected gain of attacking:
\begin{align*}
  \mathcal{G}(\hat{\rho}) =(a-1)\cdot (\mathds{P}(Y>m)\cdot \mathfrak{G})
\end{align*}
and the expected punishment as:
\begin{align*}
  \mathcal{P}(\hat{\rho}) =\mathds{P}(Y\leq m)\cdot \mathfrak{D}
\end{align*}
We can then define the expected \textit{deposit flux} per attack
attempt as the difference
$\Delta=\mathcal{P}(\hat{\rho})-\mathcal{G}(\hat{\rho})$ between the
expected punishment and the expected gain. Theorem~\ref{thm:zeroloss} shows the values for which
\blockchain yields zero-loss.
\begin{theorem}[Zero Loss Payment System]\label{thm:zeroloss}
Let 
$\rho$ be the probability of success of an attack per block, $\mathfrak{D}$ the minimum deposit per coalition expressed as a factor of the upper-bound on the gain $\mathfrak{D}=b \mathfrak{G}$, and $m$ the finalization blockdepth to return the deposit.
If $g(a,b,\rho,m)=(1-\rho^{m+1})b-(a-1)\rho^{m+1}\geq\,0$ then \blockchain implements a zero loss payment system.
\end{theorem}
\begin{proof}
Recall that the maximum gain of a successful attack is $\mathfrak{G}
\cdot (a-1)$, and the expected gain $\mathcal{G}(\hat{\rho})$ and punishment $\mathcal{P}(\hat{\rho})$ for the attackers in a
disagreement attempt are as follows:
\begin{align*} \mathcal{G}(\hat{\rho}) =&(a-1)\cdot (\mathds{P}(Y>
m)\cdot \mathfrak{G}) =(a-1)\cdot (\rho^{m+1}\cdot \mathfrak{G}),\\
\mathcal{P}(\hat{\rho}) =&\mathds{P}(Y\leq m)\cdot \mathfrak{D} =
(1-\rho^{m+1})\mathfrak{D}=(1-\rho^{m+1})b\mathfrak{G}.
  \end{align*}

Thus the deposit flux $\Delta=\mathcal{P}(\hat{\rho})-\mathcal{G}(\hat{\rho})$:
$$\Delta=\mathfrak{G}\cdot \bigg((1-\rho^{m+1})b-(a-1)\rho^{m+1}\bigg)\mathfrak{G}=g(a,b,\rho,m)\mathfrak{G}.$$

If $\Delta< 0$ then a cost of $\mathcal{G}(\hat{\rho})-\mathcal{P}(\hat{\rho})$ is incurred to the system, otherwise the punishment is enough to fund the conflicts. Thus, we explore values for which $\Delta\geq 0$. Since the gain is non-negative $\mathfrak{G}\geq 0$, it follows that $g(a,b,\rho,m)\geq 0$ for zero-loss.
\end{proof}

Setting $c=\frac{b}{a-1+b}$, we can either calculate the probability $\rho\leq c^{\frac{1}{m+1}}$ of success for an attack that \blockchain tolerates given a finalization blockdepth $m$, or a needed finalization blockdepth $m\geq \frac{\log(c)}{\log(\rho)}-1$ for a probability $\rho$ to yield zero-loss, once we fix the deposit $\mathfrak{D}$ and upper-bound the gain $\mathfrak{G}$. 

 
\paragraph{Simulating the depth and width of a coalition attack.}
Considering a deceitful ratio $\delta=\frac{f-q}{n}$, one can derive the maximum number of branches from the bound $a\leq\frac{n-(f-q)}{\ceil{2n/3}-(f-q)}$~\cite{singh2009zeno}. 
For example, for a deceitful ratio of $\delta=0.5,\, a=3$, and for a probability $\rho=0.55$, fixing the depth of the attack to $m=4$ blocks already guarantees zero-loss even if the deposit is a tenth of the maximum gain $\mathfrak{D}=\mathfrak{G}/10$, but increasing $\rho=0.9$ requires to take at least $m=28$ blocks to finish. Whereas $m$ increases polynomially with $\rho$, it increases exponentially as the deceitful ratio $\delta$ approaches the asymptotic limit $2/3$, since the number of branches also increases exponentially fast, with $m=37$ blocks for $\delta=0.6$, while $m=46$ for $\delta=0.64$ or $m=58$ for $\delta=0.66$, with $\rho=0.9$.

\begin{figure}[t]
  \hspace{-2em}
  \includegraphics[height=9em]{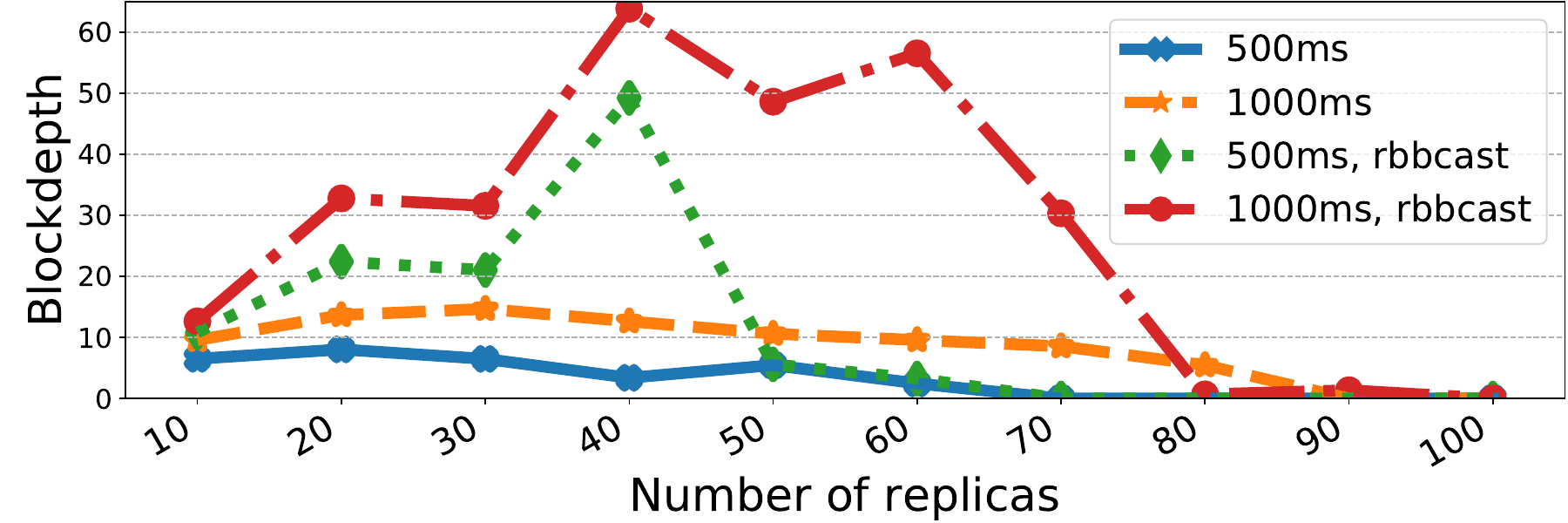}
  \caption{Minimum required finalization blockdepth $m$ to obtain zero-loss for $\mathfrak{D}=\mathfrak{G}/10$ and $f=\lceil 5n/9\rceil-1$, per number of replicas 
    `rbbcast' indicates the reliable broadcast attack.
  }
  \vspace{-1em}
  \label{fig:fig6}
\end{figure}
 

\paragraph{Experimental evaluation of the payment system.}
Taking the experimental results of~\cref{sec:expe} and based on our aforementioned theoretical analysis, Figure~\ref{fig:fig6} depicts the minimum required finalization blockdepth $m$ for a variety of uniform communication delays for $\mathfrak{D}=\mathfrak{G}/10$ and 
$f=\lceil 5n/9\rceil-1$. Again, we can see that the finalization blockdepth decreases with the number of replicas, confirming that the zero loss property scales well. 
Additionally, small uniform delays yield zero loss at smaller values of $m$, with all of them yielding $m<5$ blocks for $n>80$. 
%
%
Although omitted in the figure, our experiments showed that even for a uniform delay of 10 seconds, setting $m=50$ blocks (resp. $m=168$ blocks) still yields zero-loss in the case of a binary consensus attack (resp. reliable broadcast attack). 
Nevertheless, if the network performs normally, \blockchain will support large values of $f$, and will actually benefit from attacks (i.e., obtaining more from the deposits than is lost from the attack).

\paragraph{Discussion on probabilistic synchrony}
We assume probabilistic synchrony in this section in order to introduce a probability of failure of an attack per consensus iteration. In partial synchrony, since the committee remains static until fraudsters are identified, the adversary can successfully perform an attack with probability of success $\rho=1$. There are, however, other factors that could influence the probability $\rho$ even in partial synchrony. For example, considering a blockdepth $m\geq 1$, the implementation of a random beacon~\cite{GHM17} that replaces the committee in every iteration can decrease the probability of success of an attack. In such a case, the probability of an attack succeeding depends on the probability that the random beacon selects enough replicas of the coalition (and enough of each of the partitions of correct replicas) for $m+1$ consecutive iterations, so that the coalition is able to perform the attack for $m$ additional blocks. The design and proof of a correct random beacon that tolerates coalitions of sizes greater than $\ceil{n/3}-1$ is part of our future work.